\documentclass[12pt,english]{article}
\usepackage[T1]{fontenc}
\usepackage[latin1]{inputenc}
\usepackage{float}
\usepackage{amsmath}
\usepackage{graphicx}

\makeatletter

\providecommand{\tabularnewline}{\\}

\newcommand{\lyxaddress}[1]{
\par {\raggedright #1
\vspace{1.4em}
\noindent\par}
}

\usepackage{float}

\makeatletter

\usepackage{float}

\makeatletter

\usepackage{geometry}

\usepackage{float}


\makeatother

\makeatother

\usepackage{babel}
\makeatother

\begin{document}

\title{\textbf{\Large Sinh-Gordon Boundary TBA and Boundary Liouville Reflection
Amplitude}}

\author{Z.Bajnok $^{1}$, Chaiho Rim $^{2}$, Al. Zamolodchikov $^{3}${\normalsize }%
\thanks{On leave of absence from Institute of Theoretical and Experimental
Physics, B.Cheremushkinskaya 25, 117259 Moscow, Russia.%
}}

\maketitle
\vspace{-8cm}

\lyxaddress{\begin{flushright}
PTA/06-19\\
 RIKEN-TH-85 
\par\end{flushright}}

\vspace{6cm}

\lyxaddress{\begin{center}
\textit{$^{1}$HAS Theoretical Physics Research Group, H-1117 Budapest
Pázmány s. 1/A, Hungary}\\
 \textit{$^{2}$Department of Physics and Research Institute of Physics
and Chemistry, Chonbuk National University, Jeonju 561-756, Korea}\\
 \textit{$^{3}$Laboratoire de Physique Théorique et Astroparticules,}
UMR-5207 CNRS-UM2, \textit{Université Montpellier II, Pl.E.Bataillon,
34095 Montpellier, France} and \textit{Service de Physique Théorique,
CNRS - URA 2306, C.E.A. - Saclay F-91191, Gif-sur-Yvette, France} 
\par\end{center}}

\begin{abstract}
The ground state energy of the sinh-Gordon model defined on the strip
is studied using the boundary thermodynamic Bethe ansatz equation.
Its ultraviolet (small width of the strip) behavior is compared with
the one obtained from the boundary Liouville reflection amplitude.
The results are in perfect agreement in the allowable range of the
parameters and provide convincing support for both approaches. We
also describe how the ultraviolet limit of the effective central charge
can exceed one in the parameter range when the Liouville zero mode
forms a bound state. 
\end{abstract}

\section{Introduction}

It is often taken for granted that the short distance asymptotic of
a two dimensional relativistic field theory is described by a conformal
field theory (CFT). This conception leads to the working hypothesis
that a massive field theory can be considered as a perturbation of
its limiting CFT (CPT) by a relevant operator (or by a combination
of relevant operators) \cite{PCFT}. The corresponding (typically
dimensional) coupling constant determines the mass scale of the perturbed
model. This simple scheme holds for the most studied perturbed rational
CFT's and for certain other models like the sine-Gordon or the imaginary
coupled Toda field theories. The CPT approach also provides a systematic
description of the corrections to the ultraviolet CFT asymptotics
(with certain reservations concerning the non-analyticity in the couplings
of the vacuum expectation values, see e.g., \cite{LYcorr}). This
picture is particularly applicable for finite size effects, like the
Casimir energy \cite{Casimir}, where CPT is applied literally and
is often convergent (see e.g. \cite{TBA,Flume}). There are, however,
field theory models of different type, where the short distance asymptotic
is considerably more complicated and, up to now, no systematic description
in terms of CPT or something similar is known. Conventionally these
models can be called the {}``non-compact'' ones, since sigma-models
with non-compact target spaces are mostly of this type and reveal
the same (or sometimes more severe) peculiarities of which we're going
to discuss now.

The simplest example is the familiar sinh-Gordon model. This model
has been studied for a long time and is one of the first discovered
integrable theories \cite{VG,ArKor}. The factorized scattering amplitude
is one of the simplest possible and the complete set of form-factors
of the basic fields is known in a very explicit form \cite{Koubek,BabujanK}.
Many other characteristics such as the vacuum energy and even the
vacuum expectation values (one point correlation functions) \cite{LZ}
are known exactly (see below for a brief review). In addition, the
most general integrable boundary condition has a quite simple form
and the corresponding factorized boundary scattering admits a complete
description \cite{GZ,Ghoshal}. (This will be recapitulated briefly
in section \ref{sec:shG-scattering}).

However, it was recognized quite a while ago \cite{stair} that the
short distance asymptotic of this apparently simple model is much
more involved than the simple CPT scenario pictured above. Even if
we do not talk about the short distance behavior of the correlation
functions, which is not yet well understood even on a qualitative
footing (see \textit{e.g.}, \cite{BabujanK} for some preliminary
results), the ultraviolet behavior of the Casimir energy behaves in
quite a different way from what we're used to in CPT. The corrections
to the formal $c=1$ CFT predictions behave much softer than the usual
series in appropriate powers of the scale. It was realized \cite{AAl1}
that these leading soft corrections are mostly controlled by the so-called
Liouville reflection amplitude (LRA), a quantity of importance in
the explicit construction of the Liouville field theory (LFT) \cite{AAl1,DO}.
(For an explicit construction based on the conformal bootstrap see
\cite{JorgL}.) Although there are serious arguments to believe that
LFT plays also an important role in the description of the UV asymptotics
of other observables, including correlation functions, the finite
size settlement is probably the one where our current understanding
is the best. The relation between the UV behavior of finite size energy
and the reflection amplitudes in related non-rational CFT's, similar
to what was first argued in \cite{AAl1} for the sinh-Gordon case,
has been observed in other integrable 2D models of {}``exponential
interaction'', such as SUSY sinh-Gordon \cite{SUSy}, affine Toda
systems \cite{Toda} and the generalized sausage model \cite{sausage,fat}.

In the present publication we report a study of a somewhat different
settlement of the Casimir problem where, instead of restricting the
system to a finite circle with periodic boundary conditions, we put
it to a finite interval with integrable boundary conditions at both
sides. The ground state energy in this case is measured by means of
a modified version of the thermodynamic Bethe ansatz (TBA), the boundary
TBA (BTBA) \cite{mussardo}. The UV corrections in this case turn
out to be related to the boundary Liouville reflection amplitude (BLRA),
the boundary Liouville version of LRA. The system under consideration
turns out to be much more rich in physics than the periodic circle
one, since the boundary conditions provide enough parameters to reach
physically interesting regimes. But before turning to these interesting
topics, let us briefly remind the standard periodic Casimir effect
of the sinh-Gordon model to establish the convention.

The bulk sinh-Gordon model is defined by the Lagrangian density \begin{equation}
\mathcal{L}_{\text{sinhG}}=\frac{1}{4\pi}\left(\partial_{a}\phi\right)^{2}+2\mu\cosh\left(2b\phi\right)\label{shG}\end{equation}
 Here $\phi$ is a two-dimensional scalar field, $b$ a dimensionless
parameter and $\mu$ a dimensional coupling constant which determines
the scale of the model. In particular the physical mass $m$ of the
basic (and the only stable) particle $A$ of the model is related
to $\mu$ as \cite{masscale} \begin{equation}
\pi\mu\gamma(b^{2})=\left[\frac{m}{8\sqrt{\pi}}p^{p}(1-p)^{1-p}\Gamma\left(\frac{p}{2}\right)\Gamma\left(\frac{1-p}{2}\right)\right]^{2+2b^{2}}\label{mum}\end{equation}
 where $p$ is another convenient parameter, often used instead of
$b$\begin{equation}
p=\frac{b^{2}}{1+b^{2}}\label{p}\end{equation}
 The model is integrable and its factorized scattering theory is completely
characterized by the $AA\rightarrow AA$ scattering amplitude \begin{equation}
S(\theta)=\frac{\sinh\theta-i\sin\pi p}{\sinh\theta+i\sin\pi p}\,.\label{Sbulk}\end{equation}
 The knowledge of the scattering theory allows one to apply the TBA
to find the finite size ground state energy $E_{0}(R)$ of the model
living on a periodic circle of circumference $R$\begin{equation}
E_{0}(R)=\mathcal{E}R-\frac{m}{2\pi}\int\cosh\theta\log\left(1+e^{-\varepsilon(\theta)}\right)d\theta\label{E0}\end{equation}
 The infinite volume bulk vacuum energy $\mathcal{E}$ is also known
exactly \cite{LZ,Ebulk} \begin{equation}
\mathcal{E=}\frac{m^{2}}{8\sin\pi p}\label{Ebulk}\end{equation}
 and $\varepsilon(\theta)$ is the solution to the non-linear integral
TBA equation \begin{equation}
mR\cosh\theta=\varepsilon+\varphi*\log\left(1+e^{-\varepsilon(\theta)}\right)\label{TBA}\end{equation}
 ($*$ stands for the convolution in $\theta$). The kernel $\varphi(\theta)$
is related to the ShG scattering amplitude (\ref{Sbulk}) as \begin{equation}
\varphi(\theta)=-\frac{i}{2\pi}\frac{d}{d\theta}\log S(\theta)=\frac{1}{2\pi}\frac{4\sin\pi p\cosh\theta}{\cosh2\theta-\cos2\pi p}\label{kernel}\end{equation}
 In view of Eqs.~(\ref{mum}, \ref{Sbulk}, \ref{Ebulk}), the ShG
model possesses the weak-strong duality, $b\to1/b$ (or $p\to1-p$),
thus the analysis is restricted to $0<b^{2}<1$ (or $0<p<1/2$).

It is also convenient to introduce the {}``effective central charge''
$c_{\text{eff}}(R)$ \begin{equation}
E_{0}(R)=-\frac{\pi c_{\text{eff}}(R)}{6R}\label{ceff}\end{equation}
 instead of $E_{0}(R)$. The most important asymptotic part of the
effective central charge at $R\rightarrow0$ can be described in terms
of the {}``Liouville quantization condition'' as \begin{equation}
c_{\text{eff}}=1-24P^{2}+\text{power-like corrections in }R\label{cP}\end{equation}
 where $P$ is the solution of the transcendental equation \cite{AAl1},
\begin{equation}
\Delta_{\text{L}}(P)=\pi+4PQ\log(R/2\pi)\,.\label{dL}\end{equation}
 $\Delta_{\text{L}}(P)$ is the phase of the LRA \begin{equation}
S_{\text{L}}(P)=-\exp(i\Delta_{\text{L}}(P))\label{DL}\end{equation}
 which reads explicitly \begin{equation}
S_{\text{L}}(P)=-\left(\pi\mu\gamma(b^{2})\right)^{-2iP/b}\frac{\Gamma(1+2ibP)\Gamma(1+2ib^{-1}P)}{\Gamma(1-2ibP)\Gamma(1-2ib^{-1}P)}\,.\label{SL}\end{equation}
 Here $\mu$ is the same coupling constant as in Eq.~(\ref{shG})
and in the Liouville context is called the bulk cosmological constant.
Explicit arguments leading to the relation in Eq.~(\ref{dL}) will
be given in section \ref{sec:BLRA}, for the more complicated case
of the {}``open'' finite size effects. We mention here only that
LFT can be obtained formally as a kind of {}``reduction'' of the
Lagrangian (\ref{shG}): Neglecting one of the exponentials in the
interaction term, $2\mu\cosh\left(2b\phi\right)=\mu\exp\left(2b\phi\right)+\mu\exp\left(-2b\phi\right)$,
we are left with the familiar bulk Liouville Lagrangian \begin{equation}
\mathcal{L}_{\text{L}}=\frac{1}{4\pi}\left(\partial_{a}\phi\right)^{2}+\mu e^{2b\phi}\,.\label{LL}\end{equation}
 The former is known to define a non-rational CFT with central charge
\begin{equation}
c_{\text{L}}=1+6Q^{2}\label{cL}\end{equation}
 where $Q$ is yet another convenient parameter \begin{equation}
Q=b^{-1}+b\label{Q}\end{equation}
 and is, for historical reasons, called the {}``Liouville background
charge'' .

In what follows we are going to apply the same idea of \cite{AAl1}
to the system on a finite strip of length $R$ with appropriate {}``right''
and {}``left'' boundary conditions (henceforth referred to as {}``1''
and {}``2'', respectively), and relate the small $R$ asymptotic
of the ground state energy $E_{\text{strip}}(R)$ to the {}``boundary
Liouville reflection amplitudes'' $S_{\text{B}}(P|s_{1},s_{2})$
in \cite{FZZ}. At the same time, $E_{\text{strip}}(R)$ can be alternatively
{}``measured'' through the BTBA. General integrable boundary condition
in sinh-Gordon model contains two continuous parameters at each edge
(see section \ref{sec:shG-scattering}), so that the {}``open strip''
settlement offers, apart from the overall parameter $b$, four parameters
to play with. This makes the problem quite interesting and rich in
physical phenomena. We will start with some pedagogical reviews on
BTBA and BLRA in the first few sections and provide new results in
later sections.

The content is organized as follows. In section \ref{sec:BLRA}, we
describe briefly the boundary Liouville problem and present the explicit
expression for the boundary two-point function first given in \cite{FZZ},
which coincides with BLRA up to notations. Here the singularity structure
and the strong-weak duality of the theory are manifest in Barnes double-gamma
and double-sine function \cite{Barnes} whose definitions and useful
relations are found in the Appendix. In section \ref{sec:MSA}, semi-classical
{}``mini-superspace'' calculation \cite{BToda} is presented, which
gives an independent support to the BLRA and will feed our intuition
in later discussion.

We begin section \ref{sec:shG-scattering} with a brief survey of
the factorized boundary scattering in the boundary sinh-Gordon model
\cite{GZ,Ghoshal}. Section \ref{sec:shG-strip} is devoted to the
general formulation of the whole four parameter {}``open strip''
problem. Here we develop the usual {}``zero mode dynamics'' arguments,
which relate the UV behavior of $E_{\text{strip}}(R)$ to the {}``boundary
Liouville quantization equation'' (involving two different BLRAs,
$S_{\text{B}}(P|s_{1}^{+},s_{2}^{+})$ and $S_{\text{B}}(P|s_{1}^{-},s_{2}^{-})$).
In section \ref{sec:BTBA}, straightforward form of the related BTBA
\cite{mussardo} is presented and its analytic properties are discussed.
The standard BTBA equation, however, in a certain region of the parameters
needs manipulation of the singular behavior of the boundary fugacity
to improve the slow convergence of numerics. It is to be noted in
section \ref{sec:BTBA-singular} that BTBA is insensitive to the sign
of the boundary scattering parameters, whereas BLRA is not. This mismatch
is again due to the singular behavior of the boundary fugacity in
BTBA and appears in other BTBA problems \cite{Dorey,Saleur,Rim} as
well. BTBA is modified by introducing an additional term, relating
to the sign change of the {}``one-particle'' coupling in the boundary
state \cite{GZ}. In section \ref{sec:Numerics}, the BTBA equation
is solved numerically and is compared to the result of BLRA. Small
$R$ asymptotic from BTBA with at least one edge {}``symmetric''
(having non-singular fugacity), is found in excellent agreement with
the BLRA result when the parameters are away from the singularity
domain of BLRA. Inside the singularity domain, BRLA shows that Liouville
zero mode is not traveling anymore but trapped to form a bound state,
which in turn make the UV limit of the effective central charge to
exceed one. BTBA with both edges {}``asymmetric' (having singular
fugacity) supports perfectly the BLRA result. Section is devoted to
the analytic calculation of the boundary condition dependent UV central
charge from BTBA and. The result is found to be in agreement with
the one coming form BLRA providing another confirmation of the conjectured
relation between the UV and IR parameters. Section \ref{sec:discussion}
is the summary and discussion.

\section{Boundary Liouville reflection amplitude}

\label{sec:BLRA}

Let us consider Liouville field theory (\ref{LL}) on a strip of width
$\pi$, parametrized by the transversal ({}``space'') coordinate
$0<\sigma<\pi$ and the {}``time'' $t$ along the strip. The complex
coordinates are, as usual $\xi=\sigma+i\tau$ and $\bar{\xi}=\sigma-i\tau$.
Conformally invariant right and left boundary conditions are described
by the action \begin{equation}
\mathcal{A}=\int_{-\infty}^{\infty}d\tau\left[\int_{0}^{\pi}\left(\frac{1}{4\pi}(\partial_{a}\phi)^{2}+\mu e^{2b\phi}\right)d\sigma+M_{1}e^{b\phi}(0,\tau)+M_{2}e^{b\phi}(\pi,\tau)\right]\label{Laction}\end{equation}
 where, as before, the parameter $b$ is related to the LFT central
charge (\ref{cL}) with (\ref{Q}). $M_{1}$ and $M_{2}$ are called
the {}``right'' and {}``left'' {}``boundary cosmological constants''
as $\mu$ the bulk cosmological constant, and are conveniently parametrized
in terms of the dimensionless parameters $s_{1}$ and $s_{2}$ \cite{FZZ}
\begin{equation}
M_{1,2}=M_{0}\,\cosh(\pi bs_{1,2})\,,\quad M_{0}=\left(\frac{\mu}{\sin\pi b^{2}}\right)^{1/2}\label{s12}\end{equation}
 Since $M_{1,2}$ are real they will be parametrized as follows: When
$M_{1,2}>M_{0}$, $s_{1,2}=\tau_{1,2}\,$ with $\tau_{1,2}$ real
and positive. When $-M_{0}<M<M_{0}$, $\, s_{1,2}=ib^{-1}(1/2+b^{2}\sigma_{1,2})$
with $-b^{-2}/2<\sigma_{1,2}<b^{-2}/2$ real. When $\,\, M<-M_{0}$,
there are serious reasons to believe that BLFT is not stable anymore
and we exclude this range from the investigations.

Let us denote $\mathcal{B}$ the space of states of the LFT on the
strip. Conformal invariance entails the existence of a (single in
this case) set of generators $L_{n}$ which form the Virasoro algebra
\begin{equation}
\left[L_{m},L_{n}\right]=(m-n)L_{m+n}+\frac{c_{\text{L}}}{12}(m^{3}-m)\delta_{m+n}\,,\label{Vir}\end{equation}
 It acts on $\mathcal{B}$, splitting $\mathcal{B}$ into a set of
its highest weight representations. The Hamiltonian, the translation
generator in $\tau$, is \begin{equation}
H=-\frac{c_{\text{L}}}{24}+L_{0}\,.\label{H}\end{equation}
 The best way to understand the structure of $\mathcal{B}$ is to
take the {}``zero mode'' of the Liouville field \begin{equation}
\phi_{0}=\int_{0}^{\pi}\phi(\sigma)\frac{d\sigma}{\pi}\label{phi0}\end{equation}
 and consider the region in the configuration space where $\phi_{0}\rightarrow-\infty$.
Both the bulk and boundary interaction terms vanish in this region
and we are left with the free massless boson on the strip with free
boundary conditions on both boundaries. It is convenient, therefore,
to decompose $\phi(\sigma,\tau)$ in the oscillator modes \begin{equation}
\phi(\sigma,\tau)=\phi_{0}-2i\mathcal{P}\tau+\sum_{n\neq0}\frac{2ia_{n}}{n}e^{-n\tau}\cos\left(n\sigma\right)\label{osc}\end{equation}
 Here \begin{equation}
\mathcal{P}=-i\frac{\partial}{\partial\phi_{0}}\label{P}\end{equation}
 while the oscillators $a_{n}$ satisfy \begin{equation}
[a_{m},a_{n}]=\frac{m\delta_{m+n}}{2}\label{aa}\end{equation}
 The Virasoro generators in this {}``free field'' region are combined
as follows \begin{align}
L_{n} & =\sum_{k\neq0,n}a_{k}a_{n-k}+(2\mathcal{P}+inQ)a_{n}\ \ \ \ \ \ \ n\neq0\nonumber \\
L_{0} & =2\sum_{k>0}a_{-k}a_{k}+Q^{2}/4+\mathcal{P}^{2}\label{La}\end{align}

It is easy to argue (see \cite{AAl1}) that the conformal invariance
of the boundary theory prescribes the wave function of a primary state
$\Psi_{P}$ of dimension $\Delta_{P}=Q^{2}/4+P^{2}$ to have the following
asymptotic in the region $\phi_{0}\rightarrow-\infty$ \begin{equation}
\Psi_{P}=\left(\exp(iP\phi_{0})+S_{\text{B}}(P|s_{1},s_{2})\exp(-iP\phi_{0})\right)\left|\text{Fock vacuum}\right\rangle \;.\label{Psiass}\end{equation}
 It is the coefficient $S_{\text{B}}(P|s_{a},s_{b})$ near the {}``reflected
wave'' what is called the BLRA. Exactly as in the case of {}``bulk''
reflection \cite{AAl1}, the boundary reflection amplitude is unitary
\begin{equation}
S_{\text{B}}(P|s_{1},s_{2})=-\exp(i\Delta_{\text{B}}(P|s_{1},s_{2}))\label{DeltaB}\end{equation}
 with the phase $\Delta_{\text{B}}(P|s_{1},s_{2})$ real at real $P$.
The standard arguments of real analyticity require the analytic unitarity
\begin{equation}
S_{\text{B}}(P|s_{1},s_{2})S_{\text{B}}(-P|s_{1},s_{2})=1\,.\label{unitarity}\end{equation}
 In the usual boundary CFT language the primary state (\ref{Psiass})
is interpreted as the one created by the (juxtaposition if the right
and left boundary conditions are different) boundary operator \begin{equation}
B_{s_{1},s_{2}}^{Q/2+iP}=\exp((Q/2+iP)\phi)_{s_{1},s_{2}}\label{B}\end{equation}
 Hence, under a proper normalization the Liouville boundary reflection
amplitude is directly related to the boundary two-point function $D_{\text{B}}(\beta|s_{1},s_{2})=\left\langle B_{s_{1}s_{2}}^{\beta}(0)B_{s_{2}s_{1}}^{\beta}(1)\right\rangle $,
introduced and found in \cite{FZZ} \begin{align}
D_{\text{B}}(\beta|s_{1},s_{2}) & =\frac{\Gamma_{b}(2\beta-Q)}{\Gamma_{b}(Q-2\beta)}\times\label{DB}\\
 & \ \ \frac{\left(\pi\mu\gamma(b^{2})b^{2-2b^{2}}\right)^{(Q-2\beta)/2b}}{S_{b}\left(\beta+i\dfrac{s_{1}+s_{2}}{2}\right)S_{b}\left(\beta-i\dfrac{s_{1}+s_{2}}{2}\right)S_{b}\left(\beta+i\dfrac{s_{1}-s_{2}}{2}\right)S_{b}\left(\beta-i\dfrac{s_{1}-s_{2}}{2}\right)}\nonumber \end{align}
 Here $\Gamma_{b}(x)=\Gamma_{2}(x|b,b^{-1})$ and $S_{b}(x)=S_{2}(x|b,b^{-1})$
are the standard double-gamma and double-sine functions invented by
Barnes \cite{Barnes} (see the Appendix for a brief list of definitions
and useful relations).

The BLRA is simply the same quantity with $\beta=Q/2+iP$ \begin{align}
S_{\text{B}}(P|s_{1},s_{2}) & =D_{\text{B}}(\frac{Q}{2}+iP|s_{1},s_{2})\label{SB}\end{align}
 It is a meromorphic function of $P$ and its phase allows the power
expansion, \begin{equation}
\Delta_{\text{B}}(P|s_{1},s_{2})=\sum_{k=1}^{\infty}\Delta_{k}(s_{1},s_{2})P^{2k-1}\,.\label{DP}\end{equation}
 For practical calculations it is convenient to represent the phase
(\ref{DeltaB}) in the form \begin{equation}
\Delta_{\text{B}}(P|s_{1},s_{2})=\frac{1}{2}\Delta_{\text{L}}(P)+\Delta(P|s_{1},s_{2})\label{Delta}\end{equation}
 where $\Delta_{\text{L}}(P|s_{1},s_{2})$ is the bulk Liouville reflection
phase (\ref{DL}), while the $s$-dependent part admits a convenient
integral representation \begin{equation}
\Delta(P|s_{1},s_{2})=\int\limits _{-\infty}^{\infty}\frac{\sin(2Pt)dt}{t}\frac{\cos(s_{1}t)\cos(s_{2}t)-\cosh(bt/2)\cosh(b^{-1}t/2)\cosh(qt/2)}{\sinh(bt)\sinh(t/b)}\label{Dint}\end{equation}
 where $q=b^{-1}-b$. It is essentially a Fourier transform and is
very convenient for numerical implementation.

The states with real values of the momentum $P$ generally constitute
the continuous spectrum of {}``physical'' states \begin{equation}
\mathcal{B}=\otimes_{P\geq0}\mathcal{B}_{P}\label{Bspace}\end{equation}
 All these states are characterized by the energy $E>-1/24$. There
are, however, certain situations when additional {}``discrete''
states appear. This happens if the reflection amplitude $S_{\text{B}}(P)$
has a pole at some $P=P_{\text{b}}$ with $\Im m\, P_{\text{b}}>0$.
Then in the asymptotic (\ref{Psiass}), the incident wave is absent
and the state is localized. The pole can appear when at least one
of $M_{1}$ or $M_{2}$ is negative enough. In the $\sigma$-parametrization
the related poles in the BLRA (\ref{SB}) appear at \begin{equation}
P_{n}=i\left(\frac{\sigma_{1}+\sigma_{2}-1}{2}-n\right)b\qquad;\quad n=0,1,...,\left[\frac{(\sigma_{1}+\sigma_{2}-1)}{2}\right]\label{Pn}\end{equation}
 where $[a]$ stands for the greatest integer less than or equal to
$a$ and the boundary parameters are limited as \begin{equation}
1<\sigma_{1}+\sigma_{2}<b^{-2}\label{sigmab}\end{equation}
 (The right hand side inequality follows from the requirement of the
stability of the system $M_{a}>-M_{0}$).

For later comparisons let us quote here the semi-classical expression
for BLRA (\ref{SB}): in the limit $b\rightarrow0$, $P\rightarrow0$
and $s_{1,2}\rightarrow\infty$ while $k=P/b$ and $\sigma_{1,2}$
kept fixed \begin{equation}
S^{\text{(cl)}}(k)=\left(\frac{4\pi\mu}{b^{2}}\right)^{-ik}\frac{\Gamma(2ik)\Gamma(1/2-\sigma-ik)}{\Gamma(-2ik)\Gamma(1/2-\sigma+ik)}\label{Slim}\end{equation}
 where $\sigma=(\sigma_{1}+\sigma_{2})/2$.

\section{{}``Mini-superspace'' approximation}

\label{sec:MSA} In this section we provide a semi-classical confirmation
of the BLRA (\ref{Slim}). Let us consider the semi-classical regime
where $b\rightarrow0$ while $k=P/b$ and the boundary parameters
$\sigma_{1,2}$ are kept fixed so that \begin{equation}
M_{1,2}=-(\pi\mu)^{1/2}b\sigma_{1,2}\label{Mms}\end{equation}
 In the mini-superspace approximation one neglects all the oscillator
modes, replacing the Fock space by the vacuum state, and takes into
account only the dynamics of the {}``zero mode'' (\ref{phi0}).
The Hamiltonian (\ref{H}) is replaced by \begin{equation}
H_{\text{ms}}=-\frac{1}{24}-\frac{\partial^{2}}{\partial\phi_{0}^{2}}+\pi\mu e^{2b\phi_{0}}+(M_{1}+M_{2})e^{b\phi_{0}}\label{Hms}\end{equation}
 The corresponding eigenfunction of momentum $P=bk$ solves the second
order linear differential equation \begin{equation}
\left(-\frac{\partial}{\partial\phi_{0}^{2}}+\pi\mu e^{2b\phi_{0}}+(M_{1}+M_{2})e^{b\phi_{0}}\right)\psi(\phi_{0})=k^{2}\psi(\phi_{0})\,.\label{Dhyper}\end{equation}
 This is a degenerate hyper-geometric equation. Appropriate solution
is \begin{equation}
\psi(\phi_{0})=\left(4\pi\mu b^{-2}\right)^{-ik/2}\frac{\Gamma(1/2-ik-\sigma)}{\Gamma(-2ik)}W_{\sigma,ik}\left(2(\pi\mu)^{1/2}b\exp(b\phi_{0})\right)\label{psiW}\end{equation}
 where $\sigma=(\sigma_{1}+\sigma_{2})/2$ and \begin{equation}
W_{\lambda,\mu}(z)=\frac{z^{\mu+1/2}e^{-z/2}}{\Gamma(1/2+\mu-\lambda)}\int_{0}^{\infty}e^{-zt}t^{\mu-\lambda-1/2}(1+t)^{\mu+\lambda-1/2}dt\label{Whittaker}\end{equation}
 is the Whittaker function \cite{WW}. At $\phi_{0}\rightarrow-\infty$
\begin{equation}
\psi(\phi_{0})\sim e^{ibk\phi_{0}}-\frac{\Gamma(1+ik)\Gamma(1/2+ik)\Gamma(1/2-ik-\sigma)}{\Gamma(1-ik)\Gamma(1/2-ip)\Gamma(1/2+ik-\sigma)}\left(\frac{\pi\mu}{4b^{2}}\right)^{-ik}e^{-ibk\phi_{0}}\label{psiass}\end{equation}
 Thus the boundary reflection amplitude in this approximation reads
\begin{equation}
S^{\text{(cl)}}(p)=-\left(\frac{\pi\mu}{4b^{2}}\right)^{-ik}\frac{\Gamma(1+ik)\Gamma(1/2+ik)\Gamma(1/2-ik-\sigma)}{\Gamma(1-ik)\Gamma(1/2-ik)\Gamma(1/2+ik-\sigma)}\label{Scl}\end{equation}
 in complete agreement with the corresponding limit of the exact BLRA
(\ref{Slim}).

\section{Boundary sinh-Gordon scattering}

\label{sec:shG-scattering}In this section we analyze the sinh-Gordon
model with the Lagrangian (\ref{shG}) in the half-space $y<0$. The
boundary theory is specified by the boundary action, which in the
most general integrable case has the form \cite{GZ} \begin{equation}
A_{\text{BshG}}=\int_{y<0}\left[\frac{1}{4\pi}(\partial_{a}\phi)^{2}+2\mu\cosh(2b\phi)\right]d^{2}x+\int\left[M^{+}e^{b\phi}(0,y)+M^{-}e^{-b\phi}(0,y)\right]dy\label{AB}\end{equation}
 It will be convenient to parametrize the boundary coupling constants
$M^{\pm}$ following (\ref{s12}) through the (self-dual) parameters
$s^{+}$ and $s^{-}$ as follows \begin{equation}
M^{\pm}=M_{0}\cosh(\pi bs^{\pm})\label{Mspm}\end{equation}

Integrable boundary conditions are characterized either through integrable
boundary interactions or through factorized boundary scatterings.
The relevant amplitude of the factorized off-boundary scattering $A(\theta)B=R(\theta)A(-\theta)B$
is easily figured out from ref.\cite{GZ}. It reads as \cite{Ghoshal}
\begin{equation}
R(\theta)=R_{0}(\theta)R^{(1)}(\theta|\eta,\vartheta)\label{R}\end{equation}
 where the {}``minimal'' amplitude $R_{0}(\theta)$ is independent
of the boundary parameters \begin{equation}
R_{0}(\theta)=\frac{\sinh\left(\theta/2+i\pi/4\right)\cosh(\theta/2-i\pi p/4)\cosh(\theta/2-i\pi(1-p)/4)}{\sinh(\theta/2-i\pi/4)\cosh(\theta/2+i\pi p/4)\cosh(\theta/2+i\pi(1-p)/4)}\,.\label{R0}\end{equation}
 Note that $R_{0}(\theta)$ is singular at $\theta=i\pi/2$, which
corresponds to the emission of a zero momentum particle by the boundary
state in the crossed channel, see \cite{GZ} for the details.

The second multiplier gives the boundary parameter dependence \begin{equation}
R^{(1)}(\theta|\eta,\vartheta)=\frac{\sinh\theta-i\cosh(p\eta)}{\sinh\theta+i\cosh(p\eta)}\,\,\frac{\sinh\theta-i\cosh(p\vartheta)}{\sinh\theta+i\cosh(p\vartheta)}\label{R1}\end{equation}
 Here $\eta$ and $\vartheta$ are related to the self-dual parameter
$s^{\pm}$ \cite{FZZ,T} \begin{equation}
2b\eta=\pi(s^{+}+s^{-})\,,\qquad2b\vartheta=\pi(s^{+}-s^{-})\label{etas}\end{equation}
 Henceforth, we will call the {}``symmetric'' boundary the one with
$M^{+}=M^{-}$ (or $s^{+}=s^{-}=s$). For the symmetric boundary we
have \begin{equation}
b\eta=\pi s\,,\qquad\vartheta=0\label{symm}\end{equation}

Let us quote here the expression for the boundary energy $f(\eta,\vartheta)$
as the function of the boundary parameters $\eta$ and $\vartheta$
\begin{equation}
f(\eta,\vartheta)=\frac{m}{4\sin(\pi p)}\left(2\cosh(p\eta)+2\cosh(p\vartheta)-\sin(\pi p/2)-\cos(\pi p/2)-1\right)\label{Eb}\end{equation}
 where, as in (\ref{Ebulk}), $m$ is the mass of the fundamental
particle of the sinh-Gordon scattering theory. At the best knowledge
of the authors this expression has never been obtained rigorously.
The best way to derive it is to apply the standard relation between
the BTBA kernel and the bulk and boundary energy (see \cite{D} or
\cite{T} for details). However, strictly speaking this relation is
justified only in the case of a standard ultraviolet pattern of perturbed
rational CFT. It is the regular perturbative structure of the short
distance corrections which allows to require the cancellation of the
linear and constant terms \cite{D}. In the case of the sinh-Gordon
theory this is certainly not the case. As we mentioned in the introduction,
the ultraviolet structure is more complicated and it is not clear
for us how to ask for such cancellation against a background of much
bigger {}``soft'' corrections. Another way would be to relate the
exact one-point function of the boundary operator $\exp(b\phi)_{s,s}$
to $f(\pi sb^{-1},0)$ \cite{FZZ,T}. However this is not a derivation,
since exactly this relation has been used to figure out the relation
(\ref{etas}) between the Lagrangian and parameters of the scattering
theory. Although to our conviction there are no doubts about that
expression (\ref{Eb}) is correct, in the absence of a rigorous derivation
the analysis presented below can be considered as its important support.

\section{Sinh-Gordon on a strip}

\label{sec:shG-strip}

Now we are ready to consider the whole problem of the sinh-Gordon
model on a finite strip with two different boundary conditions at
the right and left boundaries. Let $R$ be the width of the strip.
Apart from the bulk parameters $b$ and $\mu$, we need four extra
boundary parameters $M_{1,2}^{\pm}$ to characterize the boundary
interaction at the right and left boundaries. They enter the strip
action \begin{equation}
A_{\text{strip}}=\int_{-\infty}^{\infty}L_{\text{strip}}(y)dy\label{Astrip}\end{equation}
 in the following way \begin{align}
L_{\text{strip}}(y)=\int_{0}^{R} & \left(\frac{1}{4\pi}(\partial_{a}\phi)^{2}+2\mu\cosh(2b\phi)\right)dx\label{Lstrip}\\
 & +M_{1}^{+}e^{b\phi}(0)+M_{1}^{-}e^{-b\phi}(0)+M_{2}^{+}e^{b\phi}(R)+M_{2}^{-}e^{-b\phi}(R)\,.\nonumber \end{align}
 Accordingly we need four parameters $s_{1,2}^{\pm}$ in the usual
way related to $M_{1,2}^{\pm}$\begin{equation}
M_{1,2}^{\pm}=M_{0}\cosh(\pi bs_{1,2}^{\pm})\label{Ms4}\end{equation}
 Scaling properties of the bulk and boundary fields allow to reduce
the width of the strip $R$ to $\pi$ while rendering the $R$ dependence
directly to the coupling constants. This is achieved through the rescaling
$x=(R/\pi)\sigma$ and $y=(R/\pi)\tau$. The rescaled Lagrangian reads
\begin{align}
L_{\text{strip}}(\tau) & =\int_{0}^{\pi}\left(\frac{1}{4\pi}(\partial_{a}\phi)^{2}+2\mu\left(\frac{R}{\pi}\right)^{2+2b^{2}}\cosh(2b\phi)\right)d\sigma\label{Lresc}\\
 & \ \ +\left(\frac{R}{\pi}\right)^{1+b^{2}}\left(M_{1}^{+}e^{b\phi}(0)+M_{1}^{-}e^{-b\phi}(0)+M_{2}^{+}e^{b\phi}(R)+M_{2}^{-}e^{-b\phi}(R)\right)\nonumber \end{align}
 Notice that the boundary parameters (\ref{Ms4}) are unchanged under
this rescaling.

For our present analysis it is a good idea to single out the {}``zero
mode'' (\ref{phi0}) of the field $\phi(\sigma,\tau)$ and introduce
the {}``oscillator'' operators $a_{n}$ through (\ref{osc}), (\ref{P})
and (\ref{aa}). The following Hamiltonian corresponds to (\ref{Lresc})
(compare with (\ref{H})) \begin{align}
\frac{R}{\pi}H_{\text{strip}} & =-\frac{1}{24}-\frac{\partial}{\partial\phi_{0}^{2}}+2\sum_{k>0}a_{-k}a_{k}+\mu\left(\frac{R}{\pi}\right)^{2+2b^{2}}\int_{0}^{\pi}\exp\left(2b\phi(\sigma,0)\right)d\sigma\label{Hstrip}\\
 & \ \ \ +\left(\frac{R}{\pi}\right)^{1+b^{2}}\left(M_{1}^{+}e^{b\phi}(0,0)+M_{1}^{-}e^{b\phi}(0,0)+M_{2}^{+}e^{b\phi}(\pi,0)+M_{2}^{-}e^{-b\phi}(\pi,0)\right)\,.\nonumber \end{align}
 Here the exponentials are thought as normal ordered with respect
to the operators $a_{n}$, e.g., \begin{equation}
e^{b\phi}(\pi,0)=e^{b\phi_{0}}\exp\left(-2ib\sum_{n>0}(-1)^{n}\frac{a_{-n}}{n}\right)\exp\left(2ib\sum_{n>0}\frac{a_{n}}{n}(-)^{n}\right)\,.\label{exp}\end{equation}

In the present study we are interested in the ground state energy
$E_{12}(R)$ of the strip system. In terms of the Hamiltonian (\ref{Hstrip})
it reduces to finding its lowest energy eigenvector $\Psi_{0}$\begin{equation}
H_{\text{strip}}\Psi_{0}=E_{12}(R)\Psi_{0}\label{HE0}\end{equation}
 in the space of states \begin{equation}
\mathcal{B}_{\text{strip}}=L_{2}(\phi_{0})\otimes\left(\text{Fock space of oscillators}\right)\,.\label{Bstrip}\end{equation}
 Of course in general this is a complicated infinite dimensional problem.
However, in the narrow strip (ultraviolet) asymptotic $R\rightarrow0$,
which we mostly consider in the present study, there is always a wide
{}``free region'' if $b\left|\phi_{0}\right|\ll-\log$ $\left[\left(R/\pi\right)^{b^{2}}\max(M_{1,2}^{\pm},\mu^{1/2})\right]$,
where both the bulk and the boundary interaction terms in (\ref{Hstrip})
can be neglected (see \cite{stair} for similar considerations in
the {}``closed'' geometry). Here an approximation similar to the
mini-superspace one of section \ref{sec:MSA} is rightful and \[
\Psi_{0}\sim\left(A_{+}\exp(iP\phi_{0})+A_{-}\exp(-iP\phi_{0})\right)\left|\text{Fock vacuum}\right\rangle \]
 \begin{equation}
E_{12}(R)=\frac{\pi}{R}\left(-\frac{1}{24}+P^{2}\right)\,.\label{E0P}\end{equation}
 where we introduced a ($R$ dependent) {}``momentum'' parameter
$P$. This parameter is fixed by the solution of the problem outside
the free region, requiring a usual decay of $\Psi_{0}$ at $\phi_{0}\rightarrow\pm\infty$.
In our approximation we neglect $\exp(-b\phi)$ and $\exp(-2b\phi)$
at $\phi_{0}\rightarrow\infty$, reducing the problem to the boundary
Liouville one. \begin{equation}
\frac{A_{-}}{A_{+}}=\left(\frac{R}{\pi}\right)^{-2iPQ}S_{\text{B}}(P|s_{1}^{+},s_{2}^{+})\,.\label{LQCr}\end{equation}
 Similar analysis of the interaction at $\phi_{0}\rightarrow-\infty$
results in \begin{equation}
\frac{A_{+}}{A_{-}}=\left(\frac{R}{\pi}\right)^{-2iPQ}S_{\text{B}}(P|s_{1}^{-},s_{2}^{-})\,.\label{LQCl}\end{equation}
 Eqs.~(\ref{LQCr}) together with (\ref{LQCl}) give the boundary
version of the {}``Liouville quantization condition'' (BLQC) and
for the ground state $P$ is chosen as the solution to the transcendental
equation \begin{equation}
-4PQ\log(R/\pi)+\Delta_{\text{B}}(P|s_{1}^{+},s_{2}^{+})+\Delta_{\text{B}}(P|s_{1}^{-},s_{2}^{-})=2\pi\,.\label{LQC}\end{equation}

Equations (\ref{E0P}) and (\ref{LQC}) with the boundary Liouville
reflection phases constitute our approximation. In the limit $R\rightarrow0$
the solution to $P$ is small \begin{equation}
P\sim\frac{\pi}{-2Q\log(R/\pi)}\,.\label{Plog}\end{equation}
 Therefore the smaller the value of $P$, the better is our approximation.
On general footings we expect that the leading correction to Eqs.~(\ref{E0P})
and (\ref{LQC}) are of order $R^{2bQ}$ (see \cite{shTBA} for analogous
consideration about the cylinder case). In view of (\ref{Plog}) this
means that this correction is exponentially small in $P$ and is of
the order of $\exp\left(-{b\pi}/P\right)$.

\section{Boundary TBA equation}

\label{sec:BTBA}

Formally the BTBA equation gives the strip ground state energy \cite{mussardo}
\begin{equation}
E(R)=-\frac{m}{4\pi}\int_{-\infty}^{\infty}\cosh\theta\log\left(1+\lambda_{12}(\theta)e^{-\varepsilon(\theta)}\right)d\theta\end{equation}
 where $\varepsilon$($\theta)$ is the solution to the BTBA equation
\begin{equation}
\varepsilon=2mR\cosh\theta-\varphi*L(\theta)\quad;\qquad L(\theta)=\log(1+\lambda_{12}e^{-\varepsilon})(\theta)\,.\label{BTBA}\end{equation}
 Here $*$ is ordinary convolution over the real axis of $\theta$.
The strip ground state energy is normalized to compare with the one
$E_{12}(R)$ in (\ref{E0P}) obtained using the BLQC. \begin{equation}
E_{12}(R)=E(R)+\mathcal{E}R+f_{1}+f_{2}=-\frac{\pi}{24R}c_{\text{eff}}(R)\,.\label{BTBA-energy}\end{equation}
 The quantity $\lambda_{12}(\theta)$ is called the boundary fugacity
and is given in terms of the boundary scattering amplitude in Eqs.~(\ref{R0})
and (\ref{R1}) as: \begin{equation}
\lambda_{12}(\theta)=K_{1}(-\theta)K_{2}(\theta)\end{equation}
 where $K_{1,2}(\theta)=K_{0}(\theta)k_{1,2}(\theta)$ is given in
terms of the boundary factorized scattering amplitude Eqs.~(\ref{R0},
\ref{R1}) \[
K_{0}(\theta)=R_{0}(i\pi/2-\theta)\,,\qquad k_{1,2}(\theta)=R_{1,2}^{(1)}(i\pi/2-\theta).\]
 Explicitly, the fugacity reads\begin{eqnarray}
\lambda_{12}(\theta) & = & \coth^{2}\frac{\theta}{2}\cdot\frac{\cosh\theta+\cos\frac{\pi p}{2}}{\cosh\theta-\cos\frac{\pi p}{2}}\cdot\frac{\cosh\theta+\sin\frac{\pi p}{2}}{\cosh\theta-\sin\frac{\pi p}{2}}\cdot\frac{\cosh\theta-\cos(\eta_{1}p)}{\cosh\theta+\cos(\eta_{1}p)}\nonumber \\
 & \times & \frac{\cosh\theta-\cos(\eta_{2}p)}{\cosh\theta+\cos(\eta_{2}p)}\cdot\frac{\cosh\theta-\cos(\vartheta_{1}p)}{\cosh\theta+\cos(\vartheta_{1}p)}\cdot\frac{\cosh\theta-\cos(\vartheta_{2}\, p)}{\cosh\theta+\cos(\vartheta_{2}p)}\,.\label{Bfugacity}\end{eqnarray}
 Let us note that $\lambda_{12}$ is in general singular at $\theta=0$,
which reflects the one-particle emission at the boundary \cite{GZ}.
However, this singularity vanishes if at least one of the boundaries
is chosen, say, $\vartheta_{1}=0$, to be symmetric. Then, the double
pole in $K_{0}(-\theta)K_{0}(\theta)$ is canceled with the double
zero of $k_{1}(-\theta)k_{1}(\theta)$ and the one-particle emission
disappears. In this case the numerical analysis of the BTBA equation
(\ref{BTBA}) does not show any slow convergence. Strictly speaking
this is the validity range of the original derivation of the BTBA
equation (\ref{BTBA}) in \cite{mussardo} and we are safe to apply
it only in this domain. Its extension for the case when both one-particle
boundary couplings are non-vanishing requires some care and we devote
the next section to this issue.

\section{BTBA equation with both edges asymmetric}

\label{sec:BTBA-singular}

If both boundaries are asymmetric we face with conceptual and numerical
problems. As the infrared analysis in \cite{bajnok} showed the BTBA
equation (\ref{BTBA}) describes properly the ground state energy
only for $g_{1}g_{2}>0$. Even in this case the numerical analysis
is also in trouble since even though the convolution integration in
(\ref{BTBA}) is finite, numerical evaluation is very slow in convergence.

To avoid this, one may rewrite the BTBA so that the convolution of
the singular part is analytically integrated out. One way is to put
\begin{equation}
\epsilon(\theta)=2mR\,\cosh\theta-\tau(\theta)-\varphi*L_{\ell}(\theta)\label{large_scale_TBA}\end{equation}
 where \begin{eqnarray}
L_{\ell}(\theta) & = & \log\left(\frac{1+\lambda_{12}(\theta)e^{-\epsilon(\theta)}}{1+\frac{g_{1}^{2}g_{1}^{2}e^{-2mR}}{4\sinh^{2}\theta}}\right)\\
\tau(\theta) & = & \int_{-\infty}^{\infty}\frac{d\theta'}{2\pi}\,\varphi(\theta-\theta')\,\log\left(1+\frac{g_{2}^{2}g_{2}^{2}e^{-2mR}}{4\sinh^{2}\theta}\right)\\
 & = & \frac{1}{2}\ln\left\{ \frac{\cosh\theta-\cos\pi(p+\gamma)}{\cosh\theta+\cos\pi(p+\gamma)}\,\,\,\,\frac{\cosh\theta+\cos\pi(p-\gamma)}{\cosh\theta-\cos\pi(p-\gamma)}\times\right.\nonumber \\
 &  & \left.~~~~~~~~~\frac{(\cosh2\theta-\cos2\pi(p+\gamma))(\cosh2\theta-\cos2\pi(p-\gamma))}{(\cosh2\theta-\cos2\pi p)^{2}}\right\} \,.\nonumber \end{eqnarray}
 Here $\sin\gamma\pi\equiv{|g_{1}g_{2}|e^{-mR}}/{2}\,\,$ and ${g_{1}g_{2}}=2\sqrt{\lim_{\theta\to0}\,\theta^{2}\,\,\lambda_{12}(\theta)}\,\,$
is the residue of the fugacity, identified as \begin{equation}
g_{a}=g(\eta_{a},\vartheta_{a})=2\sqrt{\cot\frac{\pi p}{4}\,\cot\frac{\pi(1-p)}{4}}\,\tan\frac{\eta_{a}p}{2}\,\tan\frac{\vartheta_{a}p}{2}\,,\qquad a=1,2\,.\label{residue-g}\end{equation}
 This form of the boundary TBA results in the energy of the form \begin{equation}
E(R)=-m\frac{\vert g_{1}g_{2}\vert}{4}e^{-mR}-\frac{m}{4\pi}\int_{-\infty}^{\infty}d\theta\,\cosh\theta\, L_{\ell}(\theta)\,.\label{E0largeL}\end{equation}
 where the one particle contribution is manifested at large distances.
For $g_{1}g_{2}>0$ it is in accordance with the boundary analog of
the Lüscher type correction \cite{bajnok} and $L_{\ell}(\theta)$
can be further expanded in even number of particle contributions,
\textit{i.e.,} powers of $e^{-2mR}$ for sufficiently large volume.

There are other choices to de-singularize the BTBA. Another useful
form \cite{KP,BLZ,Dorey2} is given as \begin{equation}
\epsilon(\theta)=2mR\,\cosh\theta-\zeta(\theta)-\varphi*L_{s}(\theta)\label{small_scale_TBA}\end{equation}
 where \begin{eqnarray*}
L_{s}(\theta) & = & \log\left(\tanh^{2}\theta+\tanh^{2}\theta\,\lambda_{12}(\theta)\, e^{-\epsilon(\theta)}\right)\\
\zeta(\theta) & = & \log{\Big(\frac{\cosh\theta+\sin\pi p}{\cosh\theta-\sin\pi p}\Big)\,\Big(\frac{\cosh2\theta+\cos2\pi p}{\cosh2\theta-\cos2\pi p}\Big)}\,.\end{eqnarray*}
 The energy in this case has the form \begin{equation}
E(R)=-\frac{m}{4\pi}\Big\{2\pi+\int_{-\infty}^{\infty}d\theta\,\cosh\theta\, L_{s}(\theta)\Big\}\,,\label{E0smallL}\end{equation}

Not only the infrared analysis suggests that the BTBA equation (\ref{BTBA})
cannot be correct for any choice of the boundary parameters if $\vartheta_{1}\cdot\vartheta_{2}\ne0$
but this can be seen also from the UV analysis: The strip ground state
energy obtained from BTBA Eq.~(\ref{BTBA-energy}) is not sensitive
to the sign change of $\vartheta$ since the boundary fugacity in
Eq.~(\ref{Bfugacity}) is not changed. On the other hand, the energy
Eq.~(\ref{E0P}) from the BLQC is sensitive to the sign change of
$\vartheta$ as we show now. Suppose we change $\vartheta_{1}$ into
$-\vartheta_{1}$, then according to the relation (\ref{etas}) $s_{1}^{+}$
turns into $s_{1}^{-}$ and vice versa so that the quantization in
Eqs.~(\ref{LQC}) reads \begin{equation}
-4PQ\log(R/\pi)+\Delta_{\text{B}}(P|s_{1}^{-},s_{2}^{+})+\Delta_{\text{B}}(P|s_{1}^{+},s_{2}^{-})=2\pi\,.\label{LQCO}\end{equation}
 This change is serious if $\vartheta_{1}\cdot\vartheta_{2}\ne0$
(see Eqs.~(\ref{Delta}, \ref{Dint})). Thus, raises a serious question
which one is correct.

This mismatch is also noted in the context of other different BTBA
problems \cite{Dorey,Saleur,Rim}. It turns out that the source of
trouble is the singular behavior of the fugacity and to cure the trouble
one needs to modify the original BTBA in Eq.~(\ref{BTBA}) when $\vartheta_{1}\cdot\vartheta_{2}\ne0$.

The correct equation can be obtained by analytical continuation in
the one-particle boundary coupling $g$ in a model-independent way:
To initiate, one notes that the double pole of the fugacity induces
a pair of zero singularity satisfying \begin{equation}
1+\lambda_{12}(\theta)e^{-\epsilon(\theta)}=0\label{singularE}\end{equation}
 on the imaginary rapidity axis. This can be easily seen at the infrared
(IR) limit. In this case putting the zero positions $\theta=iu$ and
noting $\epsilon\cong2mR\cosh\theta$, one has for Eq.~(\ref{singularE})
\[
\lambda_{12}(iu)\,\, e^{-2mR\cos u}=-1\,.\]
 The double pole structure of the fugacity results in the zeroes at,
with a good approximation \begin{equation}
u\cong\pm\frac{\vert g_{1}g_{2}\vert}{2}e^{-mR}\label{uzero}\end{equation}
 which is exponentially close to the pole at the origin.

\begin{center}
\begin{figure}[hbt]
\begin{centering}
\includegraphics[width=8cm,keepaspectratio]{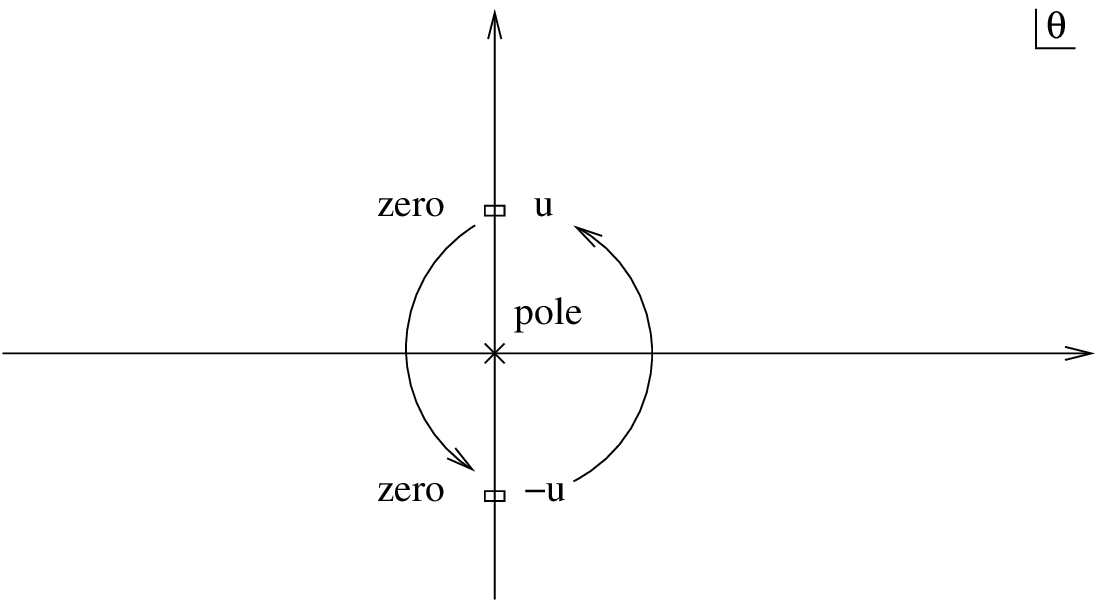} ~~~\includegraphics[width=8cm,keepaspectratio]{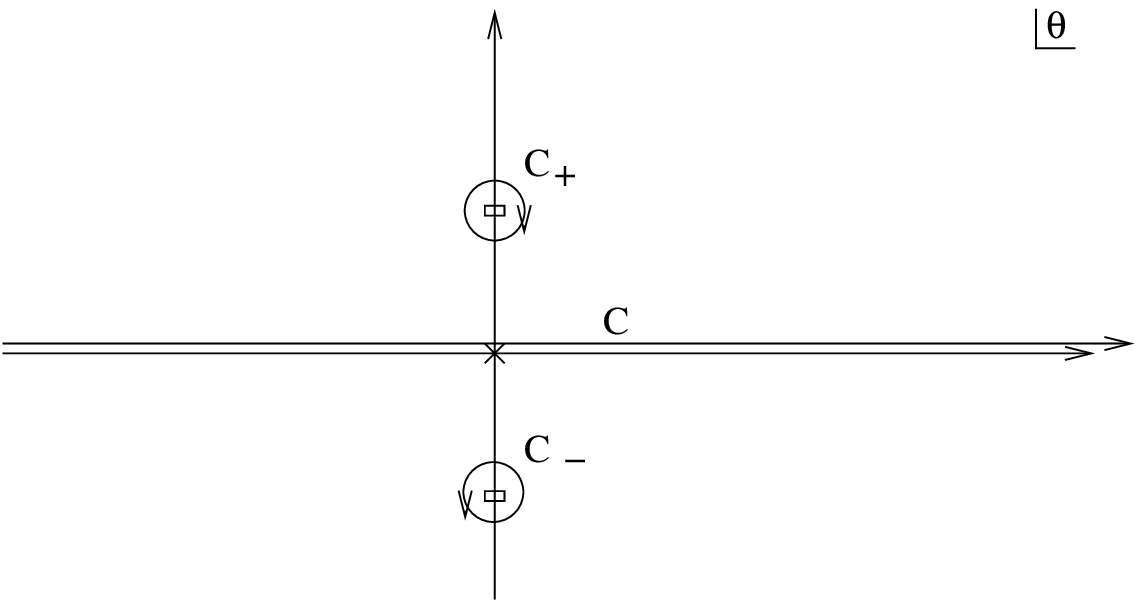} 
\par\end{centering}

\caption{Integration contour is deformed including the zero singularity positions}

\label{contour} 
\end{figure}

\par\end{center}

The convolution integral along the real axis is finite. In order to
describe the opposite sign of $g$ (or $\vartheta$) case, one analytically
continues the BTBA equation by deforming the integration contour and
picking up the zero singularity contribution as shown on Fig.~\ref{contour}.
If one integrates by part the convolution term turns the zero of the
logarithm argument to a pair of pole singularities. Finally one arrives
at the compact form of the new BTBA equation, \begin{equation}
\epsilon^{(1)}(\theta)=2mR\cosh\theta+\log\frac{S(\theta-iu)}{S(\theta+iu)}-\varphi*\log\left(1+\lambda_{12}\, e^{-\epsilon^{(1)}}\right)(\theta)\label{modifiedBTBA}\end{equation}
 where $u>0$ is the positive solution of the zero singularity \begin{equation}
\lambda_{12}(iu)\,\, e^{-\epsilon^{(1)}(iu)}=-1\,,\label{zeroes}\end{equation}
 The modified energy has the form \begin{equation}
E^{(1)}(R)=m\,\sin u-m\,\int_{-\infty}^{\infty}\frac{d\theta}{4\pi}\,\cosh\theta\,\log\left(1+\lambda_{12}(\theta)\, e^{-\epsilon^{(1)}(\theta)}\right)\,.\label{modifiedE}\end{equation}
 One may see the implication of this result easily at the IR limit.
The dominant energy becomes, with the help of $u>0$ in (\ref{uzero}),
\begin{equation}
E^{(1)}(R)=m\,\sin u-m\frac{\vert g_{1}g_{2}\vert}{4}e^{-mR}+\cdots=m\frac{\vert g_{1}g_{2}\vert}{4}e^{-mR}+\cdots\,,\label{E1largeL}\end{equation}
 which flips the sign of the IR contribution in (\ref{E0largeL}).
This result is in agreement with the boundary analog of the Lüscher
type correction \cite{bajnok}. Its confirmation in the UV region
will be provided by comparing its numerical solution with the one
obtained from the BLRA in the next section.

\section{Numerical study}

\label{sec:Numerics}

In the previous sections we presented two different expressions for
the strip ground state energy; one from BLQC Eqs.~(\ref{E0P} ,\ref{LQC})
and the other from BTBA Eqs.~(\ref{BTBA}, \ref{BTBA-energy}) or
Eqs.~(\ref{modifiedBTBA}, \ref{zeroes}, \ref{modifiedE}). These
expressions are given either as a transcendental equation or as a
nonlinear integral equation and are not easy to compare using the
analytic expression. In this section, we provide the numerical study
in a variety of parameter range.

We first note that the BLRA gives the explicit expression of $\Delta(P|s_{1},s_{2})$
in (\ref{Dint}) and $\Delta_{L}(P)$ in (\ref{DL}, \ref{SL}). On
the other hand, the ground state energy is obtained through the BLQC
Eq.~(\ref{LQC}) \begin{equation}
\Delta(P|s_{1}^{+},s_{2}^{+})+\Delta(P|s_{1}^{-},s_{2}^{-})=2\pi+4PQ\log(R/\pi)-\Delta_{L}(P)\,,\end{equation}
 which relates $P$ to the scale $R$. Thus, to compare the two different
approaches, we are enough to find the relation $R(P)$ using the BTBA.
This relation is obtained via the effective central charge through
Eq.~(\ref{BTBA-energy}), since the corresponding momentum is given
as \[
P_{\text{TBA}}=\sqrt{(1-c_{\text{eff}}(R))/24}\]
 once we use Eq.~(\ref{E0P}). Then, the Liouville boundary phase
$\Delta^{(\text{TBA})}$ from the BTBA is given as \begin{equation}
\Delta^{(\text{TBA})}(P_{\text{TBA}}|s_{1}^{+},s_{2}^{+})+\Delta^{(\text{TBA})}(P_{\text{TBA}}|s_{1}^{-},s_{2}^{-})=2\pi+4P_{\text{TBA}}Q\log\left(\frac{R(P_{\text{TBA}})}{\pi}\right)-\Delta_{L}(P_{\text{TBA}})\,.\label{BTBA-DL}\end{equation}
 Thus, the numerical check is to compare (\ref{BTBA-DL}) with the
analytic expression (\ref{Dint}). It is noted that the bulk expression
$\Delta_{L}(P)$ in (\ref{DL}, \ref{SL}) turns out to be in excellent
agreement with the TBA result even up to the order of $P\cong1$ (see
some of the results in Ref.~\cite{AAl1}).

\begin{figure}[H]
\begin{centering}
\includegraphics[width=15cm,height=9cm]{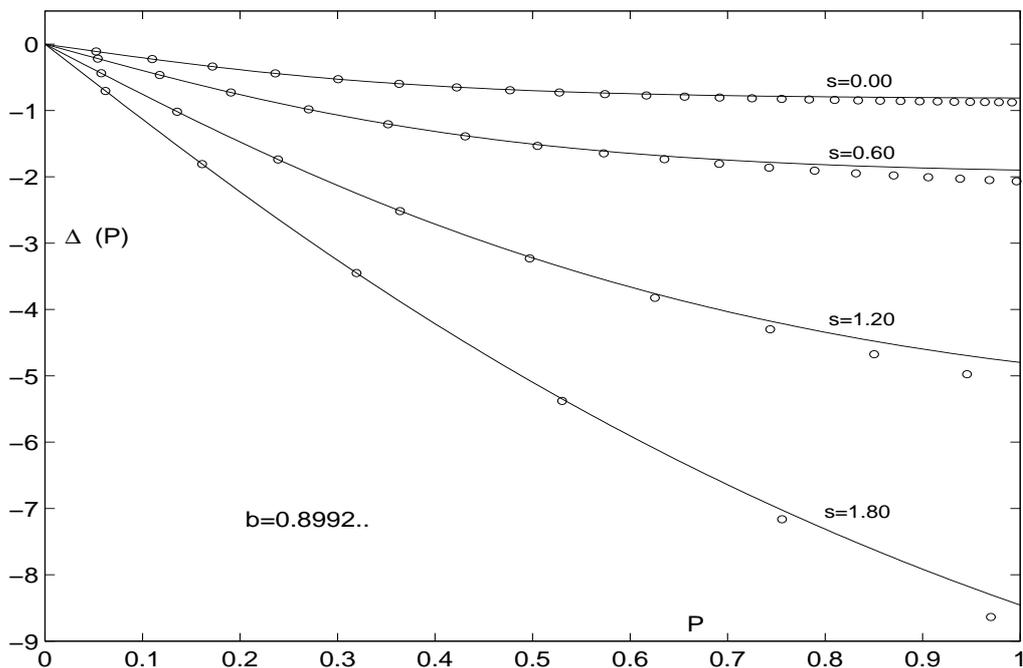} 
\par\end{centering}

\caption{$\Delta(P|s,s)$ \textit{v.s.\/} $P$ with real $s$: Solid lines
refer to the Liouville expression, while circles represent BTBA result
($b^{2}=0.8086$ is taken). \label{fig:realpara}}

\end{figure}

\subsection{Symmetric case}

\begin{figure}[H]
\begin{centering}
\includegraphics[width=15cm,height=10cm]{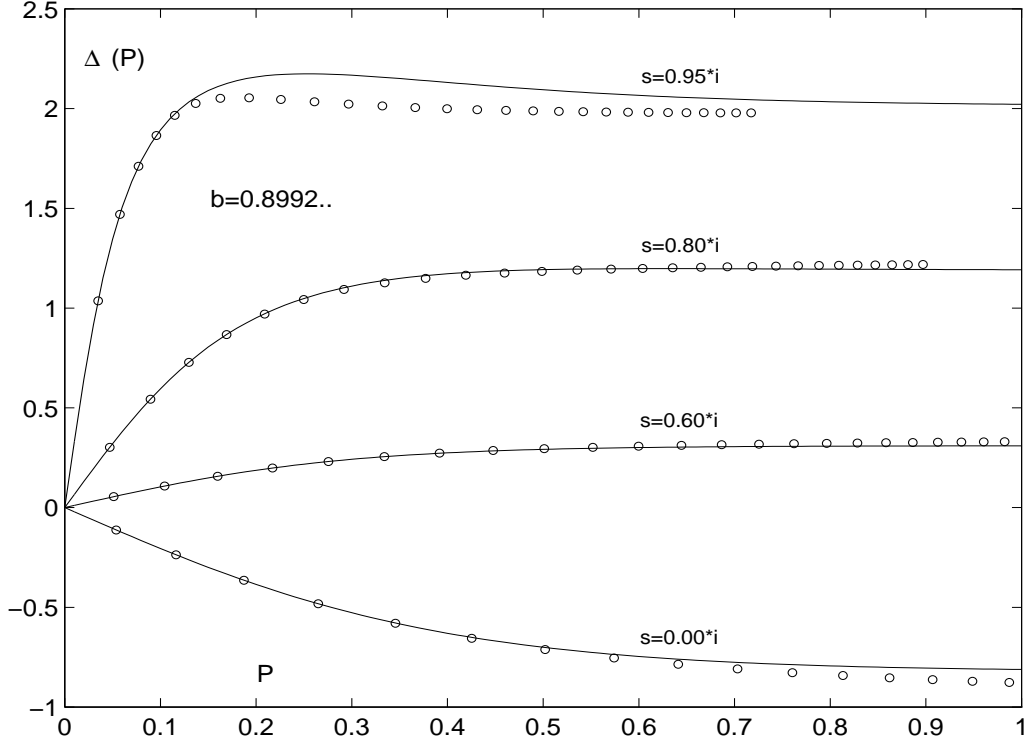} 
\par\end{centering}

\caption{$\Delta(P|s,s)$ \textit{v.s.\/} $P$ with imaginary $s$: Solid
lines refer to the Liouville expression, while circles represent BTBA
result ($b^{2}=0.8086$ is taken). \label{fig:imagpara}}

\end{figure}

We first restrict ourselves to the case with at least one edge being
symmetric, \textit{i.e.,} $s_{1}^{+}=s_{1}^{-}=s_{1}$ or $\vartheta_{1}=0$
since in this parameter range, we can avoid the singular behavior
of the boundary fugacity $\lambda_{12}(\theta)$. The simplest case
is when both boundaries are symmetric so that the right edge also
has $s_{2}^{+}=s_{2}^{-}=s_{2}$ or $\vartheta_{2}=0$ but $s_{2}$
is not necessarily the same as $s_{1}$. Here we can use that $b\eta_{a}=\pi s_{a}$
($a=1,2$). In this case, Eq.~(\ref{BTBA-DL}) simplifies to \begin{equation}
\Delta^{(\text{TBA})}(P_{\text{TBA}}|s_{1},s_{2})=\pi+2P_{\text{TBA}}Q\log\left(\frac{R(P_{\text{TBA}})}{\pi}\right)-\frac{1}{2}\Delta_{L}(P_{\text{TBA}})\,.\end{equation}
 In Fig.~\ref{fig:realpara} the numerical results for $\Delta^{(\text{TBA})}$
are compared with the analytic boundary Liouville expression for the
case of two identical boundaries $s_{1}=s_{2}=s$ at $b^{2}=0.8086$
with purely real values of $s>0$. Another plot is presented for imaginary
values of $s$ in Fig.~\ref{fig:imagpara}.

\begin{table}[H]
 \centering $\begin{tabular}{|l|l|l|}
\hline  $P$  &  $\Delta(P,s,s)$  &  $\Delta^{\text{TBA}}(P,s,s)$\\
\hline 0.06651092551935  &  0.41762001693342  &  0.41762001693341\\
\hline 0.07278412235910  &  0.45306073688754  &  0.45306073688672\\
\hline 0.08039322566412  &  0.49476603900486  &  0.49476603897984\\
\hline 0.08983119563236  &  0.54444716822153  &  0.54444716749005\\
\hline 0.10187859377520  &  0.60443320709236  &  0.60443318659429\\
\hline 0.11785948248982  &  0.67790492787209  &  0.67790438878486\\
\hline 0.14024177942679  &  0.76910412419994  &  0.76909136428327\\
\hline 0.17429836730472  &  0.88301300229112  &  0.88276474709427\\
\hline 0.23395539043946  &  1.02163214862749  &  1.01846195305733\\
\hline 0.36739025335041  &  1.15834702793920  &  1.14485912927470\\
\hline 0.74325569097388  &  1.19642293140014  &  1.21070661865468\\\hline \end{tabular}$

\caption{Result for $\Delta(P|s,s)$ \textit{v.s.\/} $P$ when $s=0.80i$
and $b^{2}=0.8086$. \label{table1}}

\end{table}

At $P$ smaller than $0.15$ the numerical agreement is impressively
good, as it is illustrated in Table~\ref{table1} for the example
of $s=0.80i$.

The agreement is also quite good up to $P\sim1$ where the values
of $R$ are already well bigger than the correlation length $m^{-1}$
and we expect the power corrections in $R$ to come into play. This
can be explained by the fact that after the contributions of the boundary
and bulk vacuum energy are added in Eq.~(\ref{BTBA-energy}), the
power corrections to $c_{\text{eff}}$ begin with a rather high power
of $R$ (indeed, they are expected to be $\sim R^{2+2b^{2}}$ at $b<1$).

\begin{figure}[H]
\begin{centering}
\includegraphics[width=15cm,height=10cm]{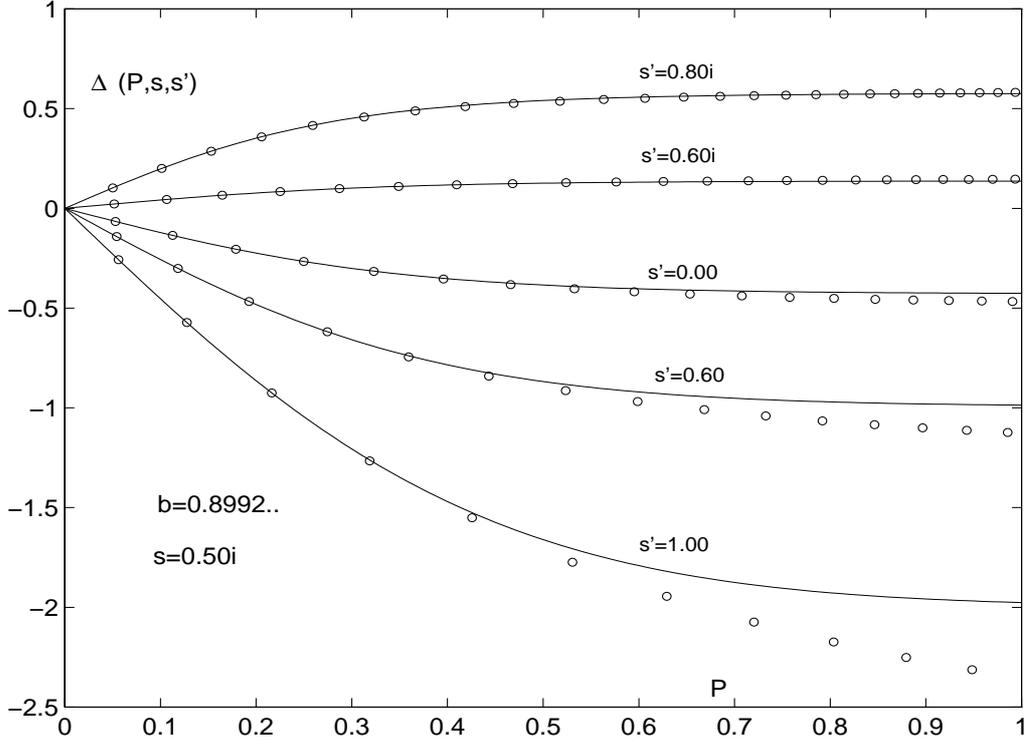} 
\par\end{centering}

\caption{$\Delta(P|s,s')$ \textit{v.s.\/} $P$: Solid lines refer to the
Liouville expression, while circles represent BTBA result ($s=0.5i$
and $b^{2}=0.8086$).\label{fig3}}

\end{figure}

Different boundaries with $s_{1}=s$ and $s_{2}=s'$, allow to measure
the Liouville reflection phase $\Delta(P|s,s')$. In Fig.~\ref{fig3}
the results are presented for $s=0.5i$ and $b^{2}=0.8086$. %

\subsection{Discrete mode's case}

\begin{figure}
\begin{centering}
\includegraphics[width=12cm,height=7cm]{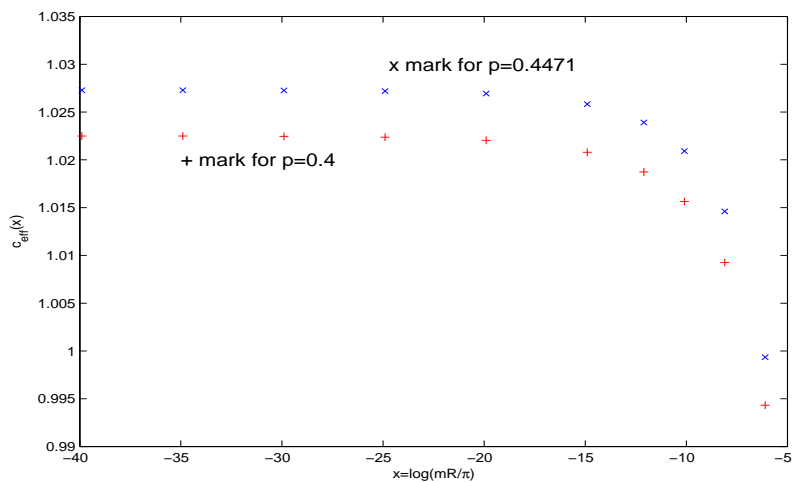} 
\par\end{centering}

\caption{$c_{\text{eff}}$ \textit{v.s.\/} $\log R\/$\,: +-marks are given
for $p=0.4$ (BLRA predicts $c_{\text{eff}}(0)=1.0225$) and x-marks
are given for $p=0.4471$ (BLRA predicts $c_{\text{eff}}(0)=1.02729025$).\label{fig:ceff-bound} }

\end{figure}

As far as one of the edges is symmetric, the two approaches, BTBA
and BLRA, are in good agreement. However, as $s_{1}$ and $s_{2}$
approach to the critical value $\text{Im}(s_{1}+s_{2})=Q$, where
$\Delta(P|s,s)$ becomes singular, the agreement fails except at a
small region of $P$, which is seen in Fig.~\ref{fig:imagpara} when
$s_{1}=s_{2}=0.95i$ and $b^{2}=0.8086$ (in this case the actual
critical value is $s=1.00565i$). In fact, there is a parameter range
where $\text{Im}(s_{1}+s_{2})$ exceeds the critical value $Q$ so
that the BLRA has the pole at imaginary value of $P$ Eq.~(\ref{Pn})
and at the same time, Im($2b\eta_{a})<\pi Q\,\,$ and Im($2b\vartheta_{a})<\pi Q$
for $a=1,2$ so that there is no boundary bound state in the IR boundary
scattering theory. This range is given by the following conditions
satisfied simultaneously: \begin{equation}
1<\sigma_{1}^{\pm}+\sigma_{2}^{\pm}<\frac{1}{b^{2}}\,,\qquad\sigma_{1}^{+}+\sigma_{1}^{-}<1\,,\qquad\sigma_{2}^{+}+\sigma_{2}^{-}<1\,.\label{sigma-bound}\end{equation}
 In this region, the UV limiting value of $c_{\text{eff}}$ exceeds
1 as shown in Fig.~\ref{fig:ceff-bound}. Then, a question arises:
How is $c_{\textrm{eff}}$ in Eq.~(\ref{BTBA-energy}) related to
the one Eq.~(\ref{E0P}) from BLQC?

\begin{figure}
\begin{centering}
\includegraphics[width=10cm]{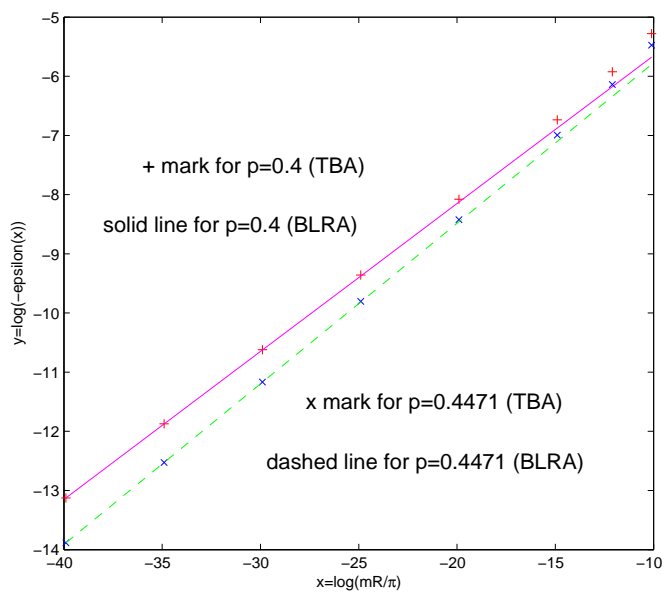}\label{discrete} 
\par\end{centering}

\caption{$\log(-\epsilon(R))$ is plotted against $\log(\frac{mR}{\pi})$ for
parameters $p=0.4$ and $p=0.4471$. Solid and dashed lines represent
results from BLRA, while the marks $+$ and $\times$ correspond to
the BTBA values. }

\end{figure}

\begin{table}
\begin{centering}
\begin{tabular}{|c|c|c|}
\hline 
$\log(\frac{mR}{\pi})$  & $c_{\mathrm{eff}}(R)$  & $\log(-\epsilon(R))$ BTBA\tabularnewline
\hline
\hline 
-39.9  & 1.022497070241340  & -13.125612063335536\tabularnewline
\hline 
-34.9  & 1.022489761509328  & -11.874296674954588 \tabularnewline
\hline 
-29.9  & 1.022464134631873  & -10.620393772536636 \tabularnewline
\hline 
-24.9  & 1.022373521967475  & -9.359087748545065 \tabularnewline
\hline 
-19.9  & 1.022045862147555  & -8.077079274139289 \tabularnewline
\hline 
-14.9  & 1.020790366172398  & -6.736968924454419\tabularnewline
\hline 
-12.1  & 1.018739648008074  & -5.923623092706940 \tabularnewline
\hline 
-10.1  & 1.015637398421726  & -5.279891692845977 \tabularnewline
\hline
\end{tabular}
\par\end{centering}

\caption{$c_{\mathrm{eff}}(R)$ and $\log(-\epsilon(R))$ obtained from BTBA
for various values of $\log(\frac{mR}{\pi})$ at $p=0.4$: }

\label{p04} 
\end{table}

\begin{table}
\begin{centering}
\begin{tabular}{|c|c|c|}
\hline 
$\log(\frac{mR}{\pi})$  & $c_{{\rm \mathrm{eff}}}(R)$  & $\log(-\epsilon(R))$ BTBA\tabularnewline
\hline
\hline 
-39.9  & 1.027288737814910  & -13.883502224634348\tabularnewline
\hline 
-34.9  & 1.027284375181948  & -12.526342850041335\tabularnewline
\hline 
-29.9  & 1.027267391791812  & -11.167552023824138 \tabularnewline
\hline 
-24.9  & 1.027200890055835  & -9.803579618174346 \tabularnewline
\hline 
-19.9  & 1.026935938305069  & -8.423633052788508 \tabularnewline
\hline 
-14.9  & 1.025827041718239  & -6.995003046786642\tabularnewline
\hline 
-12.1  & 1.023913306574199  & -6.139867067007655\tabularnewline
\hline 
-10.1  & 1.020921463329385  & -5.473674443519166\tabularnewline
\hline
\end{tabular}
\par\end{centering}

\caption{$c_{{\rm \mathrm{eff}}}(R)$ and $\log(-\epsilon(R))$ obtained from
BTBA for various values of $\log(\frac{mR}{\pi})$ at $p=0.4471$ }

\label{p044} 
\end{table}

At first sight, this question seems not to make any sense since the
parameters simply violate the convergence of $\Delta(P|s_{1},s_{2})$.
As discussed in section \ref{sec:BLRA}, this corresponds to the case
when the primary operator is not reflected at the Liouville potential
wall but is trapped inside as a bound state, which can be easily given
in the semi-classical approximation in section \ref{sec:MSA}. This
happens when $M_{1,2}^{\pm}$ is sufficiently negative, but not too,
so still maintaining the stability of the system. This suggests that
the Hilbert space has the discrete spectrum as well as the continuous
one (see also \cite{Jorg1,Jorg2}). As a consequence, the bound state
energy with Hamiltonian Eq.~(\ref{Hstrip}) is given as \[
E_{12}^{(n\pm)}(R)=\frac{\pi}{R}\left(-\frac{1}{24}+P_{n\pm}^{2}\right)\]
 where $P_{n\pm}$ is given in (\ref{Pn}). \begin{equation}
P_{n\pm}=i\left(\frac{\sigma_{1}^{\pm}+\sigma_{2}^{\pm}-1}{2}-n_{\pm}\right)b\,,\qquad n_{\pm}=0,1,...,\left[\frac{(\sigma_{1}^{\pm}+\sigma_{2}^{\pm}-1)}{2}\right]\label{Pn-pm}\end{equation}
 This will give the UV limiting value of the effective central charges
greater than 1. \begin{equation}
c_{\textrm{eff }}^{n\pm}(0)=1+24|P_{n\pm}|^{2}>1\,.\label{eq:cpm}\end{equation}
 The $c_{\text{eff}}(R)$ is given in Table~\ref{p04} and Table~\ref{p044}
and is plotted in figure \ref{fig:ceff-bound} corresponding to the
parameters $\sigma_{1}^{+}=\sigma_{2}^{-}=17.5/40,\sigma_{2}^{+}=25.5/40,\sigma_{2}^{-}=9.5/40$,
which satisfies the bound in Eq.~(\ref{sigma-bound}). Eq.~(\ref{eq:cpm})
predicts $c_{\text{eff}}^{+}(0)=1.0225$ when $p=0.4$, and $c_{\text{eff}}^{+}(0)=1.02729025\cdots$
when $p=0.4471$, which agree with the BTBA results. Since BTBA is
derived from the saddle point of the partition function it always
reproduces the lowest energy state.

We checked this UV limiting value of the effective central charge
from BTBA for various ranges of the parameters and found a complete
agreement. Interestingly, the leading corrections to the effective
central charge $c_{\mathrm{eff}}(R)$ are no longer logarithmic in
the volume but powerlike. They are not of perturbative origin, however,
but are governed by the analytical continuation of the BLQC as we
now show. For this we rewrite the BLQC in the exponentiated form\begin{equation}
S_{B}(P\vert s_{1}^{+},s_{2}^{+})S_{B}(P\vert s_{1}^{-},s_{2}^{-})=\left(\frac{R}{\pi}\right)^{4iPQ}\label{eq:ACBLQC}\end{equation}
 The discrete mode corresponds to the pole singularity of one of the
BLRA say $S_{B}(P\vert s_{1}^{+},s_{2}^{+})$. As the volume decreases
$P$ gets close to the pole at $P_{n+}$as $P=P_{n+}+i\epsilon$,
so we approximate the BLRA in the neighborhood as \[
S_{B}(P\vert s_{1}^{+},s_{2}^{+})=i\frac{G(s_{1}^{+},s_{2}^{+})}{P-P_{n+}}+\dots=\frac{G(s_{1}^{+},s_{2}^{+})}{\epsilon}+\dots\]
 For small enough $R$ we determine $\epsilon$ from (\ref{eq:ACBLQC})\begin{equation}
\epsilon(R)=G(s_{1}^{+},s_{2}^{+})S_{B}(P_{n+}\vert s_{1}^{-},s_{2}^{-})\left(\frac{R}{\pi}\right)^{4\vert P_{n+}\vert Q}\label{eq:EpsBLQC}\end{equation}
 The corresponding central charge can be written as $c_{\mathrm{eff}}(R)=1+24\vert P\vert^{2}=1+24\Big(\vert P_{n+}\vert+\epsilon(R)\Big)^{2}$,
which gives \begin{equation}
\epsilon(R)=\sqrt{\frac{c_{\mathrm{eff}}(R)-1}{24}}-\vert P_{n+}\vert\label{eq:Epsceff}\end{equation}

In the same spirit we compared the BTBA with the exact BLRA we can
calculate $\epsilon(R)$ from BTBA and compare with the expression
(\ref{eq:EpsBLQC}). The data of $\log(-\epsilon(R))$ from BTBA for
various values of $\log(\frac{mR}{\pi})$ at $p=0.4$ and $p=0.4471$
is given in Table~\ref{p04} and Table~\ref{p044} and the log-plot
is given in Figure \ref{discrete}. The expected slope is $4\vert P_{n+}\vert Q$
which is $0.25$ for $p=0.4$ (and $0.27129$ for $p=0.4471$). The
fitted value using the lower 4 points is $0.2510$ for $p=0.4$ (and
$0.2719$ for $p=0.4471$). Small $R$ results give more accurate
slope and one can see the complete agreement. It proves the correctness
of the BLRA not only for real but also for imaginary values of $P$.

\subsection{Asymmetric case}

Next, we are considering the case when both of the edges are asymmetric.
The BLQC is given as a combination of different boundary Liouville
phases. Lets us suppose that $\vartheta_{1}\cdot\vartheta_{2}>0$,
thus we have \[
\Delta(P|s_{1}^{+},s_{2}^{+})+\Delta(P|s_{1}^{-},s_{2}^{-})=2\pi+4PQ\log(R/\pi)-\Delta_{L}(P)\,,\]
 then if we switch one of the sign of $\vartheta$'s so that $\vartheta_{1}\cdot\vartheta_{2}<0$,
we have \[
\Delta(P|s_{1}^{+},s_{2}^{-})+\Delta(P|s_{1}^{-},s_{2}^{+})=2\pi+4PQ\log(R/\pi)-\Delta_{L}(P)\,.\]
 Since individual phases are confirmed already using the case with
at least one boundary symmetric, this settlement looks not to provide
any new information to the Liouville phase. Nevertheless, this combination
is important to check the correctness of the analytically continued
BTBA.

\begin{figure}[H]
\begin{centering}
\includegraphics[width=12cm,height=6cm]{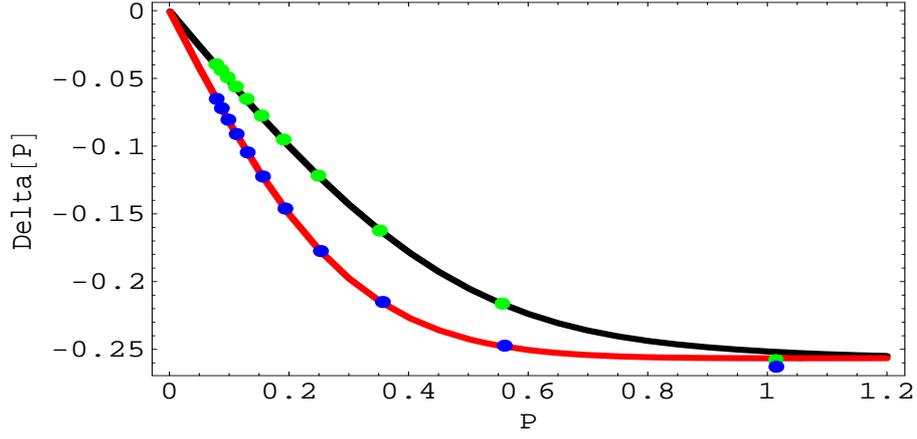} 
\par\end{centering}

\caption{Plot of $\Delta_{S}(P)$ (upper black line and green dots, for $\vartheta_{1}\cdot\vartheta_{2}>0$)
and $\Delta_{A}(P)$ (lower red line and blue dots, for $\vartheta_{1}\cdot\vartheta_{2}>0$)
for the asymmetric boundaries with $b^{2}=0.8086$ and $s_{1}^{+}=s_{2}^{+}=i2b/3$
and $s_{1}^{-}=s_{2}^{-}=0$: Solid lines refer to the Liouville expression
and circles to the BTBA and modified BTBA result. \label{deltansym}}

\end{figure}

The numerical results are in perfect agreement with the exact Liouville
amplitude if the improved version of BTBA Eqs.~(\ref{large_scale_TBA},
\ref{E0largeL}) or Eqs.~(\ref{small_scale_TBA}, \ref{E0smallL})
is applied when $\vartheta_{1}\cdot\vartheta_{2}>0$, and the modified
BTBA Eqs.~(\ref{modifiedBTBA}, \ref{zeroes}, \ref{modifiedE})
is applied when $\vartheta_{1}\cdot\vartheta_{2}<0$. This result
is plotted in Fig.~\ref{deltansym} when $s_{1}^{-}=s_{2}^{-}=0$,
for simplicity. In this case, the phase combination becomes \begin{align*}
\Delta_{A}(P)=\frac{\Delta(P|s_{1}^{+},0)+\Delta(P|0,s_{2}^{+})}{2} & =\pi+2PQ\log\left(\frac{R}{\pi}\right)-\frac{\Delta_{L}(P)}{2}\quad\textrm{for}\,\,\vartheta_{1}\cdot\vartheta_{2}<0\\
\Delta_{S}(P)=\frac{\Delta(P|s_{1}^{+},s_{2}^{+})+\Delta(P|0,0)}{2} & =\pi+2PQ\log\left(\frac{R}{\pi}\right)-\frac{\Delta_{L}(P)}{2}\quad\textrm{for}\,\,\vartheta_{1}\cdot\vartheta_{2}>0\,.\end{align*}
 Other convincing numerical checks are presented in Table (\ref{table:delta_org})
and Table (\ref{table:delta-added}). %

\begin{table}[h]
\centering $\begin{tabular}{|l|l|l|}
\hline  $P$  &  $\Delta_{S}(P)$  &  $\Delta_{S}^{\text{TBA}}(P)$\\
\hline 0.077653999816446  &  -0.039568923372291  &  -0.039568923376408 \\
\hline 0.086205185262923  &  -0.043897517771010  &  -0.043897517774418 \\
\hline 0.096861805449936  &  -0.049277259792747  &  -0.049277259789521 \\
\hline 0.110502372967937  &  -0.056134917955759  &  -0.056134917808053 \\
\hline 0.128563844490344  &  -0.065154112175326  &  -0.065154108313167 \\
\hline 0.153553843986511  &  -0.077485060440331  &  -0.077484967209008 \\
\hline 0.190215910849849  &  -0.095154635137037  &  -0.095152687636201 \\
\hline 0.248446714967505  &  -0.121766433889740  &  -0.121735638152979 \\
\hline 0.351196987049572  &  -0.162583158000073  &  -0.162317800581419 \\
\hline 0.556084052286610  &  -0.216307776233087  &  -0.216228485970792 \\
\hline 1.013529248959518  &  -0.252173074348906  &  -0.257965815998832 \\\hline \end{tabular}$

\caption{Result for $\Delta_{S}(P)$ \textit{v.s.\/} $P$ when $b^{2}=0.8086$
and $s_{1}^{+}=s_{2}^{+}=i2b/3$ and $s_{1}^{-}=s_{2}^{-}=0$. \label{table:delta_org}}

\end{table}

\begin{table}[h]
\centering $\begin{tabular}{|l|l|l|}
\hline  $P$  &  $\Delta_{A}(P)$  &  $\Delta_{A}^{\text{TBA}}(P)$\\
\hline 0.078277417261084  &  -0.065157660050461  &  -0.065157660061779 \\
\hline 0.086965071942987  &  -0.072051773672923  &  -0.072051773682382 \\
\hline 0.097806408331036  &  -0.080511089260293  &  -0.080511089203979 \\
\hline 0.111704015386026  &  -0.091098590825966  &  -0.091098589162912 \\
\hline 0.130133954494199  &  -0.104648454630161  &  -0.104648417456781 \\
\hline 0.155666785945641  &  -0.122400071889728  &  -0.122399351214232 \\
\hline 0.193135842394272  &  -0.146101177532259  &  -0.146089809923920 \\
\hline 0.252485260446018  &  -0.177599477043236  &  -0.177471319390988 \\
\hline 0.356217516450171  &  -0.215816508521331  &  -0.215038322905223 \\
\hline 0.560049812334757  &  -0.247974299790094  &  -0.247287310445203 \\
\hline 1.014342456385944  &  -0.256659409495335  &  -0.262840890495731 \\\hline \end{tabular}$

\caption{Result for $\Delta_{A}(P)$ \textit{v.s.\/} $P$ when $b^{2}=0.8086$
and $s_{1}^{+}=s_{2}^{+}=i2b/3$ and $s_{1}^{-}=s_{2}^{-}=0$. \label{table:delta-added}}

\end{table}

\section{Calculation of the UV central charge from BTBA\label{sec:ceff}}

In this section we analyze the UV behavior of the BTBA equation (\ref{BTBA}).
We are able to describe the leading small volume behavior of the central
charge analytically using the idea developed in \cite{stair}: We
expand the Fourier transform of the BTBA kernel \begin{equation}
\frac{\varphi(k)}{2\pi}=\int e^{ik\theta}\varphi(\theta)\frac{d\theta}{2\pi}=\frac{\cosh(\frac{\pi k(1-2p)}{2})}{\cosh(\frac{\pi k}{2})}=\sum_{n=0}^{\infty}(-1)^{n}\frac{\varphi_{2n}}{(2n)!}k^{2n}=1-\alpha^{2}\frac{k^{2}}{2}+\dots\label{eq:phiexpand}\end{equation}
 where $\alpha=\pi\sqrt{p(1-p)}=\frac{\pi}{Q}$ and write the BTBA
equation (\ref{BTBA}) for the rescaled functions $\theta\to\theta+x$
with $x=\log mR$ in the form of an infinite order ordinary differential
equation as \begin{equation}
e^{\theta}+e^{2x-\theta}+\log(1-e^{-L(\theta)})=\log\lambda_{12}(\theta)-\sum_{n=1}^{\infty}\frac{\varphi_{2n}}{(2n)!}L^{(2n)}(\theta)\label{eq:infode}\end{equation}
 We approximate this equation in various rapidity domains in different
manners. Since $L(\theta)$ is even we restrict the considerations
to the $\theta<0$ region. Furthermore we are interested in the $R\to0\,,\,\, x\to-\infty$
limit that is we neglect the $e^{2x-\theta}$ term, keeping in mind
that we have the same contribution from this term in the $\theta>0$
domain. We distinguish three rapidity regions as follows: $x\approx\theta$,
$x\ll\theta\ll0$ and $\theta\approx0$. If $\theta\approx x$ the
fugacity term is not relevant but we have to keep all the derivatives
of $L$. If $x\ll\theta\ll0$ we can additionally neglect the $e^{\theta}$
term together with higher derivatives of $L$. In this domain, which
we call the plateaux domain, $L$ is large and positive so we can
approximate the BTBA equation as \begin{equation}
\frac{\alpha^{2}}{2}L^{''}(\theta)+e^{-L(\theta)}=0\label{eq:infodeappr}\end{equation}
 The corresponding solution is\begin{equation}
L(\theta)=\log\frac{\sin^{2}\lambda(\theta-a)}{\lambda^{2}\alpha^{2}}\label{eq:Ltreal}\end{equation}
 with two arbitrary parameters, $\lambda,a$, which can be fixed from
the boundary conditions. The corresponding central charge is \begin{equation}
c_{\textrm{eff }}(x)=1-\frac{6\lambda^{2}\alpha^{2}}{\pi^{2}}+\dots\label{eq:ceffalpha}\end{equation}
 As we decrease the volume, $R\to0$, the plateaux region becomes
larger and larger and the approximation is better and better. So in
this way we describe the leading correction to the central charge.

The boundary condition at $x$ is provided by the kinetic term $e^{\theta}$
as $L(x)=0$. In contrast, the boundary condition at the origin is
determined by the boundary fugacity. If \begin{equation}
(\log\lambda_{12})(k=0)=\int_{-\infty}^{\infty}d\theta\log(\lambda_{12})<0\label{eq:lambdacond}\end{equation}
 then we can demand the $L(0)=-\infty$ boundary condition. This results
in \[
\lambda=\frac{\pi}{(\alpha-x)}\quad;\qquad a=0\]
 and gives the leading UV behavior of the central charge \[
c_{\textrm{eff }}=1-\frac{6\alpha^{2}}{(x-\alpha)^{2}}+\dots=1-\frac{6\pi^{2}}{Q^{2}x^{2}}+\dots\]
 which is in accord with the result of BLRA (\ref{Plog}).

In the opposite case when $(\log\lambda_{12})(k=0)>0$, the parameter
$\lambda$ turns out to be imaginary $\lambda=i\kappa$ and we have
to fit the parameters of the following function:

\begin{equation}
L(\theta)=\log\frac{\sinh^{2}\kappa(\theta-a)}{\kappa^{2}\alpha^{2}}\label{eq:Limag}\end{equation}
 Demanding $L(x)=0$ we have $a=x-\alpha$. The variable $\kappa$
is determined from the boundary condition at the origin, which is
provided by the boundary fugacity. Clearly $L$ behaves as $L\propto-2\kappa\theta$
at the middle of the plateaux region, and we will determine the value
of $\kappa$ by comparing to the solution around the origin $\theta\approx0$.
Here the kinetic term can be neglected (but not the fugacity) and
we arrive at the equation\begin{equation}
\epsilon=-\varphi*L\quad,\qquad L(\theta)=\log\left(1+\lambda_{12}(\theta)e^{-\epsilon(\theta)}\right)\label{eq:BTBAlin}\end{equation}
 If we additionally suppose that $\epsilon$ is large negative, which
follows from $L\propto-2\kappa\theta$ we arrive at the equation \[
\log\lambda_{12}-L=-\varphi*L\]
 which can be solved by Fourier transformation \begin{eqnarray*}
L(k) & = & \frac{(\log\lambda_{12})(k)}{1-\varphi(k)}\\
 & = & \frac{2\pi}{k}\frac{\left[\sinh\frac{k\pi}{2}+\sinh\frac{k\pi}{2}(1-p)+\sinh\frac{k\pi}{2}p-2\sinh\frac{k\pi}{2}(1-\frac{2\eta p}{\pi})-2\sinh\frac{k\pi}{2}(1-\frac{2\vartheta p}{\pi})\right]}{2\sinh\frac{\pi k}{2}(1-p)\sinh\frac{\pi k}{2}p}\end{eqnarray*}
 where we used the formula \[
\log(\left[x\right]_{\frac{i\pi}{2}-\theta}\left[x\right]_{\frac{i\pi}{2}+\theta})=\int\frac{dk}{2\pi}e^{-ik\theta}\frac{2\pi}{k}\frac{\sinh\frac{k\pi}{2}(2-x)}{\cosh\frac{k\pi}{2}}\]
 valid for $1<x<3$ and can be extended for $x<1$ via the relation
$\left[x\right]_{\frac{i\pi}{2}-\theta}\left[x\right]_{\frac{i\pi}{2}+\theta}=\left[2-x\right]_{\frac{i\pi}{2}-\theta}\left[2-x\right]_{\frac{i\pi}{2}+\theta}$.
We also put $\eta_{1}=\eta_{2}=\eta$ and $\vartheta_{1}=\vartheta_{2}=\vartheta$.
If in the Fourier transform we have a singular term around the origin
as $-\frac{\Lambda}{k^{2}}$ then in its inverse Fourier transform
we have a behavior as $-\frac{\Lambda}{2}\theta$. So by inspecting
the singularity structure around the origin we can extract that \begin{equation}
\kappa=\frac{1-\frac{2p}{\pi}(\eta+\vartheta)}{p(1-p)}\label{eq:kappaeta}\end{equation}
 which no longer depends on $x$ and increases the central charge.
The central charge calculated from $\kappa$ yields \begin{equation}
c_{\textrm{eff }}(x)=1+\frac{6\kappa^{2}\alpha^{2}}{\pi^{2}}+\dots=1+6\frac{(1-\frac{2p}{\pi}(\eta+\theta))^{2}}{p(1-p)}+\dots>1\label{eq:cgeqone}\end{equation}
 which agrees with the result coming form the BLRA (\ref{eq:cpm})
when the Liouville zero mode is trapped in the Liouville potential.
This gives a convincing analytical support for the UV-IR relation
(\ref{etas}) and shows the correctness of both the BLRA and the BTBA.

\section{Discussion}

\label{sec:discussion}

We have analyzed the ground-state energy of the sinh-Gordon theory
defined on the strip subject to integrable boundary conditions in
two complementary ways using BTBA and BLRA.

BTBA, being a nonlinear integral equation, systematically sums up
the finite size corrections to the infinite volume ground-state energy
by taking into account the information on the semi-infinite boundary
scattering theory. As a consequence it is formulated in terms of the
boundary reflection factors and is reliable in the IR regime. In the
case of the boundary scattering theories corresponding to perturbed
rational BCFTs the careful analysis of the UV limit of BTBA allows
the determination of the central charge together with the perturbative
power-like corrections. We have shown in the paper that, in contrast
to this usual behaviour, the UV limit of the boundary sinh-Gordon
theory is governed by a non-rational BCFT: the BLFT. The ground-state
energy acquires soft (logarithmic) corrections in the volume determined
by the BLRA, the most important quantity in the bootstrap solution
of the BLFT. This approach is valid in the UV regime and describes
the ground-state energy in terms of the parameters of the Lagrangian.

As a first step we solved numerically the BTBA and compared with the
predictions coming form the BLRA. In general, we found a convincing
evidence of the correctness of both approaches. In particular, we
checked the previously conjectured relationship between the IR and
UV parameters (\ref{etas}) and confirmed the predictions of BLRA.
Then we used the results of BLRA to check the analytically continued
BTBA.

The semi-classical picture, provided in the paper, suggested the existence
of a discrete part of the Hilbert space, which corresponds to the
case when the Liouville zero mode is trapped in the boundary Liouville
potential. We confirmed the adequacy of this picture at the quantum
level by numerically calculating the effective central charge, which
exceeds one in this case. By adopting a method to compute analytically
the leading behaviour of the UV central charge we were able to derive
its value exactly. This provides another support for both the UV-IR
relation and BLRA.

Besides confirming the validity of BLRA, which is a widely used quantity
in 2D quantum gravity, we provided evidence for the discrete part
of the Hilbert space. It would be interesting to analyze further its
consequences.

The way we performed the analytical continuation in the one-particle
boundary coupling in BTBA makes possible to apply the result directly
to other theories, like boundary Toda theories, where the integrable
boundary conditions form a discrete set and there is no room for playing
with any continuous parameter. As a consequence, the reflection factors
computed from the boundary bootstrap principle in the IR can be compared
via the modified BTBA to the parameters of the ATFT valid in the UV.
This will help to find the sofar unrevealed correspondence between
the two sets of integrable boundary conditions.

In \cite{Jorg3} the finite volume description of the sinh-Gordon
model originating from an integrable lattice realization was analyzed.
It would be nice to perform a similar analysis for the boundary sinh-Gordon
theory and explore the analogue of the trapped Liouville mode on the
lattice.

Finally, we note that similarly to the analysis of the bulk staircase
model \cite{stair} its boundary version can be investigated further,
along the line of \cite{LSS}, in order to understand better its UV
limiting theory. 

\vspace{1cm}

\textbf{Acknowledgments.} This project was initiated during the focus
program of APCTP, 2005 and supported in part by the Joint Research
Project under The KOSEF-HAS (F01-2005-000-10282-0)(R\&B), by the Center
for Quantum Spacetime (CQUeST) of Sogang University with grant number
R11-2005-021(R), and by the EGIDE project (Al.Z). One of the authors
(Al.Z) thanks the Kawai Theoretical Laboratory at RIKEN, especially
H.Kawai and T.Tada, for hospitality and stimulating scientific atmosphere
during his visit while this study was finalized. ZB was supported
by a Bolyai Scholarship, OTKA K60040 and the EC network {}``Superstring''.

\vspace{1cm}

\textbf{Note added} This work was almost finished last year while
two of the authors visited Montpellier and appeared in a final form
a month ago. After a communication through the phone with Alyosha
on October 15th, we were waiting for his last comment on section 9
which would be done by Thursday (18th). But very sadly and unexpectedly,
we are told he passed away on that night. Remembering how excited
Alyosha was about the result of this non-compact field theory, we
put this draft on the web. We have enjoyed life and discussions with
Alyosha.

\section{Double gamma and double sine}

The double gamma function $\Gamma_{b}(x)$ was introduced by Barnes
\cite{Barnes} through the analytic continuation in $z$ of the double
zeta-series \begin{equation}
\log\zeta_{b}(x,z)=\sum_{m,n=0}^{\infty}\left(x+mb+nb^{-1}\right)^{-z}\label{eq:logzetab}\end{equation}
 convergent if $z>2$ (we suppose that $\Re e\, b>0$). The analytic
continuation can be achieved by the following integral representation
\begin{equation}
\log\zeta_{b}(x,z)=\frac{\Gamma(1-z)}{2\pi i}\int_{C}\frac{e^{-xt}(-t)^{z}}{(1-e^{-bt})(1-e^{-t/b})}\frac{dt}{t}\label{eq:logzetabintrep}\end{equation}
 where the contour $C$ goes from $+\infty$ to $+\infty$ encircling
the brunch cut of $(-t)^{z}$ counterclockwise. The double gamma function
is defined as \begin{equation}
\Gamma_{b}(x)=\left.\frac{\partial}{\partial z}\zeta_{b}(x,z)\right|_{z=0}\end{equation}
 Like ordinary gamma function, $\Gamma_{b}(x)$ is a meromorphic function
with no zeros and simple poles located at $x=-mb-nb^{-1}$ with $(m,n)$
a pair of non-negative integers. All these poles are inside the {}``wedge''
\[
\left|\arg x\right|>\pi-\arg b\]
 (we imply here that $\Im m\, b\geq0$), which for real $b$ shrinks
to the negative part of the real axis. Outside the {}``wedge'' it
can be represented as the integral which follows directly from (\ref{eq:logzetabintrep})
\begin{equation}
\log\Gamma_{b}(x)=\frac{C_{\text{E}}}{2}\left(\frac{(Q-2x)^{2}}{4}-\frac{b^{2}+b^{-2}}{12}\right)+\frac{1}{2\pi i}\int_{C}\frac{e^{-tx}\log(-t)}{(1-e^{-bt})(1-e^{-t/b})}\frac{dt}{t}\end{equation}
 where $C_{\text{E}}=-\Gamma^{\prime}(1)$ is the Euler's constant.

The following dual shift relations are readily derived e.g. from the
integral representation \begin{align}
\Gamma_{b}(x+b) & =\frac{(2\pi)^{1/2}}{b^{1/2-bx}\Gamma(bx)}\Gamma_{b}(x)\label{shiftG}\\
\Gamma_{b}(x+1/b) & =\frac{(2\pi)^{1/2}}{\Gamma(x/b)b^{x/b-1/2}}\Gamma_{b}(x)\nonumber \end{align}

At large $\left|x\right|$ outside the wedge the Stirling asymptotic
expansion applies \begin{equation}
\log\Gamma_{b}(x)\sim\left(\frac{Q^{2}-2}{24}-\frac{(Q/2-x)^{2}}{2}\right)\log x+\frac{3x^{2}}{4}-\frac{Qx}{2}+\sum_{k=1}^{\infty}\frac{(k-1)!d_{k+2}(Q)}{x^{k}}\end{equation}
 Here $d_{k}(Q)$ are polynomials in $Q$ defined as \begin{equation}
d_{k}(Q)=\sum_{n=0}^{k}\frac{(-)^{n}B_{n}B_{k-n}}{n!(k-n)!}b^{2n-k}\end{equation}
 and $B_{n}$ are usual Bernoulli numbers. One of the effective numerical
algorithms is to use several times one of the shift relations (whichever
is more convenient) to render the argument to a region where the Stirling
formula with a reasonable number of asymptotic terms is effective.
For moderate values of $x$ the following form also gives quite accurate
results \begin{align*}
\log\Gamma_{b}(x)= & \left(\frac{Q^{2}-2}{24}-\frac{(Q/2-x)^{2}}{2}\right)\log x+\frac{C_{\text{E}}}{2}\left(\frac{(Q-2x)^{2}}{4}-\frac{b^{2}+b^{-2}}{12}\right)\\
 & \quad-\frac{1}{\pi}\int_{-\infty}^{\infty}\frac{e^{t(1+iz)^{2}}\log((1+iz)^{2})}{(1-e^{-b(1+iz)^{2}/x})(1-e^{-b^{-1}(1+iz)^{2}/x})}\frac{dz}{(1-iz)}\end{align*}
 There is also a convenient line integral representation \begin{equation}
\log\Gamma_{b}(x)=\int\limits _{0}^{\infty}\frac{dt}{t}\left[\frac{e^{-xt}}{(1-e^{-bt})(1-e^{-t/b})}-\frac{1}{t^{2}}-\frac{Q/2-x}{t}-\left(\frac{(x-Q/2)^{2}}{2}-\frac{b^{2}+b^{-2}}{24}\right)e^{-t}\right]\end{equation}

The diperiodic sine $S_{b}(x)$ (aka as the Barnes double sine function)
is related to $\Gamma_{b}(x)$ as \[
S_{b}(x)=\frac{\Gamma_{b}(x)}{\Gamma_{b}(Q-x)}\]
 In the strip $0<\Re e\, x<Q$ it allows the following integral:\begin{equation}
\log S_{b}(x)=\frac{1}{2}\int\limits _{-\infty}^{\infty}\frac{dt}{t}\left[\frac{\sinh(Q-2x)t}{2\sinh(bt)\sinh(t/b)}-\frac{(Q/2-x)}{t}\right]\label{intS}\end{equation}
 Being a Fourier transform this representation is convenient for numerical
calculations. Outside the strip of convergence $S_{b}(x)$ is restored
via one of two dual shift relations (which probably inspired the name
of the function) \begin{align}
S_{b}(x+b) & =2\sin(\pi bx)S_{b}(x)\label{shift}\\
S_{b}(x+1/b) & =2\sin(\pi x/b)S_{b}(x)\nonumber \end{align}
 It is a meromorphic function of $x$ with poles at $x=-mb-nb^{-1}$
with $m$ and $n$ non-negative integers. The only zeros at $x=Q+mb+nb^{-1}$
are predicted by the {}``unitarity relation'' \begin{equation}
S_{b}(x)S_{b}(Q-x)=1\end{equation}
 which is a direct consequence of (\ref{intS}). The following argument
doubling relation is useful to arrive at (\ref{DB}) in the main body
of the paper \begin{equation}
S_{b}(2x)=S_{b}(x)S_{b}(x+b/2)S_{b}(x+b^{-1}/2)S_{b}(x+Q/2)\end{equation}

\end{document}